\documentclass[fleqn,usenatbib]{mnras}

\usepackage{newtxtext,newtxmath}

\usepackage[T1]{fontenc}

\DeclareRobustCommand{\VAN}[3]{#2}
\let\VANthebibliography\thebibliography
\def\thebibliography{\DeclareRobustCommand{\VAN}[3]{##3}\VANthebibliography}

\usepackage{graphicx}	
\usepackage{amsmath}	

\usepackage{caption}
\usepackage{subcaption}
\usepackage{booktabs}
\usepackage[section]{placeins}
\usepackage{float}
\usepackage{tablefootnote}

\title[Retrograde resonances in planetary systems]{A numerical study of fourth and fifth order retrograde mean motion resonances in planetary systems}

\author[A. C. Signor et al.]{
Alan Cefali Signor,$^{1}$\thanks{E-mail: alan.cefali@unesp.br}
Gabriel Antonio Caritá,$^{1,2}$
Maria Helena Moreira Morais,$^{1}$\thanks{E-mail: helena.morais@unesp.br}
\\
$^{1}$Instituto de Geociências e Ciências Exatas, Universidade Estadual Paulista (UNESP), Av. 24-A, 1515, 13506-900 Rio Claro, SP, Brazil \\
$^{2}$ Divisão de Mecânica Espacial e Controle, INPE,
12227-310 São José dos Campos, SP, Brazil}

\date{Accepted 08 Jun 2022. Received 11 May 2022; in original form ZZZ}

\pubyear{2022}

\begin{document}
\label{firstpage}
\pagerange{\pageref{firstpage}--\pageref{lastpage}}
\maketitle

\begin{abstract}
We present a numerical study on the stability of all fourth and fifth order retrograde mean motion resonances ($1/3, 3/1, 1/4, 4/1, 2/3, 3/2$) in the 3-body problem composed of a solar mass star, a Jupiter mass planet and an additional body with zero mass (elliptic restricted problem) or masses corresponding to either Neptune, Saturn or Jupiter (planetary problem). The fixed point families exist in all cases and are identified through libration of all resonant angles simultaneously. In addition, configurations with libration of a single resonant angle were also observed. Our results for the elliptic restricted 3-body problem are in agreement with previous studies of retrograde periodic orbits, but we also observe new families not previously reported. Our results regarding stable resonant retrograde configurations in the planetary 3-body problem could be applicable to extra-solar systems.

\end{abstract}




\section{Introduction}

In the last decade, several numerical and analytical studies in the framework of the circular restricted 3-body problem provide a solid base regarding the knowledge of the  resonant dynamics of asteroids with high inclination orbits, in particular the case of retrograde configurations (inclination close to 180 degrees with respect to the ecliptic) \citep{morais2013retrograde,namouni2015resonance,morais2016numerical}. These studies led to the identification of Centaurs in retrograde resonances with Jupiter and Saturn \citep{morais2013asteroids}, and (514107) Ka`epaoka`awela in the co-orbital resonance with Jupiter \citep{wiegert2017retrograde,morais2017reckless,namouni2018coorbital}. Recently, families of periodic orbits related to retrograde resonances in the circular and elliptic restricted 3-body problems were computed by \cite{kotoulas2020retrograde, kotoulas2020planar,kotoulas2022phase}.

The simulations by \cite{malmberg2011effects} showed that close encounters between stars can lead to planet exchange and thus may generate planetary systems with high  relative inclinations. The possibility of counter-revolving configurations in extra-solar systems (relative inclination between the planets near 180 degrees) was investigated by \cite{gayon2008retrograde,gayon2009fitting}. These authors showed that fitting radial velocity curves for retrograde instead of prograde configurations could, in some cases, lead to smaller residuals. Therefore, it is possible that some of the known extra-solar systems have counter-revolving planets. 

In \cite{paper12022}, we studied the stability of  planetary systems in the 1/1, 1/2 and 2/1 retrograde resonances. These systems were composed by a solar mass star, a prograde planet with the mass of Jupiter and a retrograde planet with zero mass (elliptic restricted 3-body problem) or masses corresponding to either Neptune, Saturn or Jupiter (planetary 3-body problem). In the current article, we extend this study to the retrograde resonances 1/3, 3/1, 1/4, 4/1, 2/3 and 3/2. In Sect.~2 we describe our numerical methods and present results for the retrograde resonances 1/3, 3/1, 1/4, 4/1, 2/3 and 3/2 in the elliptic restricted and planetary cases (ER3BP and 3BP). In Sect.~3 we discuss our results and present the conclusions.

\section{Numerical study of fourth and fifth retrograde mean motion resonances}

The system considered in the simulations is composed of a solar mass star and two bodies orbiting the star in opposite directions, the prograde one in counter-clock wise motion and the retrograde one in clockwise motion. We use a notation $p/-q$ to represent a retrograde resonance with mean motion ratio $p/q$.  The  astrocentric orbital elements are: $a$ (semi-major axis), $e$ (eccentricity), $I$ (orbital inclination), $\omega$ (argument of pericenter), $\Omega$ (longitude of the ascending node), $M$ (mean anomaly), $T$ (orbital period), $\varpi$ (longitude of pericenter) $\lambda$ (mean longitude). Variables without subscript refer to the retrograde body, while variables with subscript $p$ refer to the prograde body. The prograde body has unitary semi-major axis ($a_p = 1.0$) and a mass of $0.001 \, M_{\odot}$, thus it always interact gravitationally with other bodies. The semi-major axis of the retrograde body is fixed at the nominal resonance location $a = (q/p)^{2/3}$. The retrograde body has zero mass in the elliptic restricted 3-body problem (ER3BP), and in the planetary 3-body problem it has mass equal to either Neptune ($0.00005419 \, M_{\odot}$), Saturn ($0.0002857 \, M_{\odot}$) or Jupiter ($0.001 \, M_{\odot}$). Recalling that a retrograde $p/-q$ mean motion resonance has order $p+q$ \citep{morais2013retrograde}, we investigate the stable configuration of planar systems in all fourth and fifth order retrograde mean motion resonances (1/-3, 3/-1, 1/-4, 4/-1, 2/-3, 3/-2).

The numerical integrations were performed using REBOUND with the adaptive step integrator Bulirsch-Stoer \citep{rein2012rebound}. The integration was stopped when the distance to the star was large than $10 \, a_p$ (escape) or when the distance between the two bodies was smaller than the sum of their radii (collision).

Assuming a counter-clockwise reference frame, the longitudes are defined in the orbital plane for the prograde planet (counter-clockwise motion) and are measured in the direction of the object's orbital motion, so $\lambda_p = \varpi_p + M_p$ where $\varpi_p = \omega_p + \Omega_p$. For the retrograde planet (clockwise motion) the longitudes are measured against the direction of the object's orbital motion, hence $\lambda = \varpi - M$ where $\varpi = \Omega - \omega$. This is also the convention used within REBOUND.

The resonant maps indicate in which regions the libration of the resonant angle of the circular restricted 3-body problem occurs; for a $p/-q$ mean motion resonance this resonant angle is $\phi_0 = -q \lambda - p \lambda_p + (p+q) \varpi$ \citep{morais2013retrograde}. The regions with libration of both $\phi_0$ and $\Delta \varpi = \varpi - \varpi_p$ indicate a fixed point family of the resonant problem, these are identified by a white symbol. The color bar of the resonant maps indicates the semi-amplitude libration of $\phi_0$ (its maximum variation around the resonant center value). To facilitate interpretation of the figures, we relate a color to each resonant argument different from $\phi_0$.
Regions of libration of one of these angles were represented by symbols in their respective colors. These colors are also used in the figures with the evolution of the orbital elements. We identify the two type of families that may occur in the resonant maps when both $\phi_0$ and $\Delta \varpi$, or one of the others resonant arguments librate (around either 0 or $\pi$), with semi-amplitude less than $\pi/4$. 

In our numerical simulations the planets are in a quasi-coplanar configuration, with initial inclination of the retrograde planet  $I = 179.99^\circ$. If we considered the strictly 2D case, we would also observe vertical unstable families but these should not occur in realistic systems. We set the longitude of the nodes to be zero ($\Omega = \Omega_p = 0)$

\begin{figure*}
         \includegraphics[width=\textwidth]{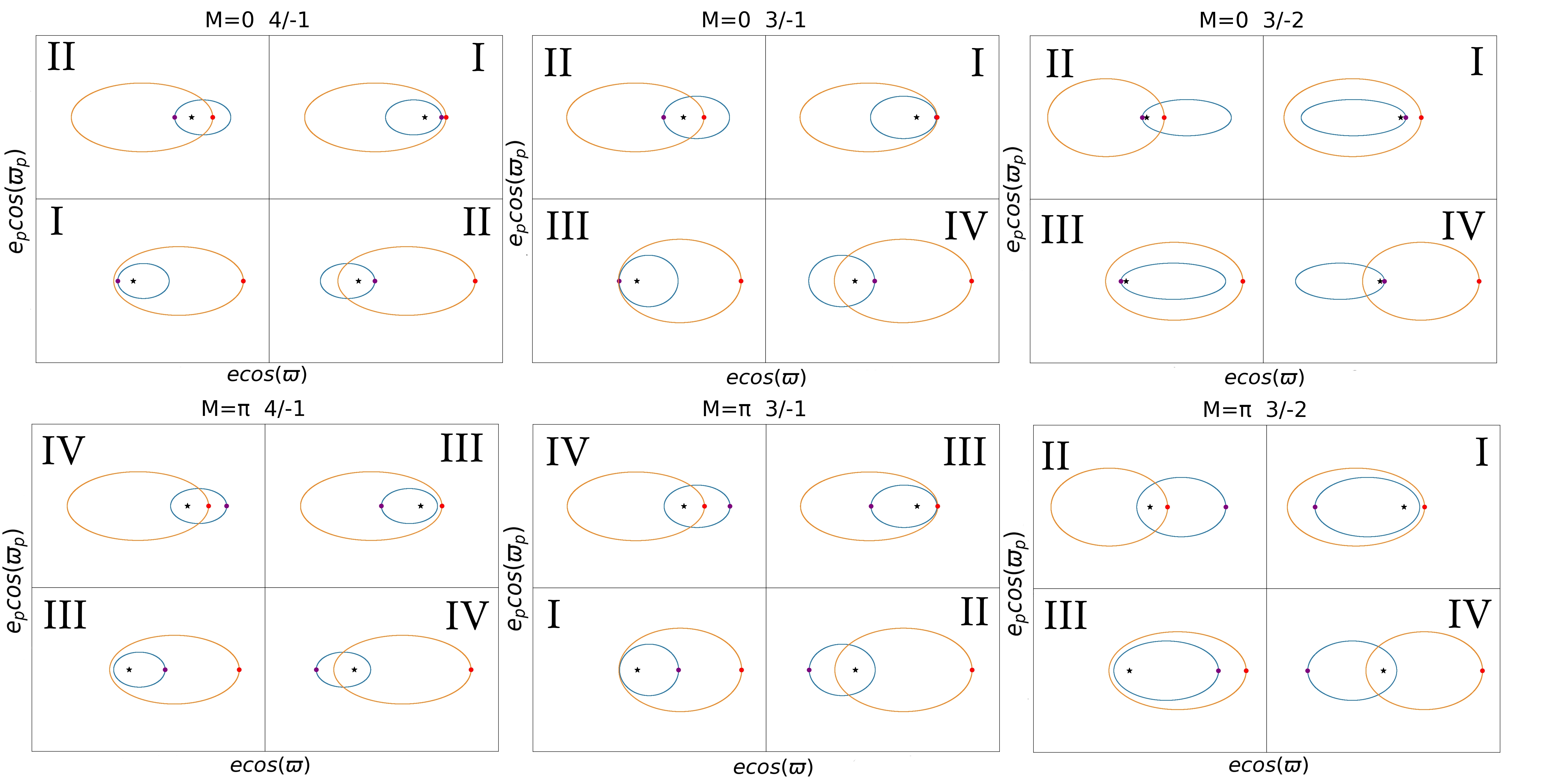}
         \caption{Configurations for the internal resonances displayed in 4 quadrants illustrating the orientations of the pericenters and apocenters at initial  mean anomalies  $M = 0$ and $M = \pi$. The Roman numerals indicate pairing of the initial configurations which, due to the commensurability between the orbital periods, are equivalent with a time-lag of half a period of the external object (4/-1 and 3/-1). For 3/-2 the configurations they are equivalent with a time-lag of an entire period.}
         \label{inertialinner0}
 \end{figure*}

\begin{figure*}
         \includegraphics[width=\textwidth]{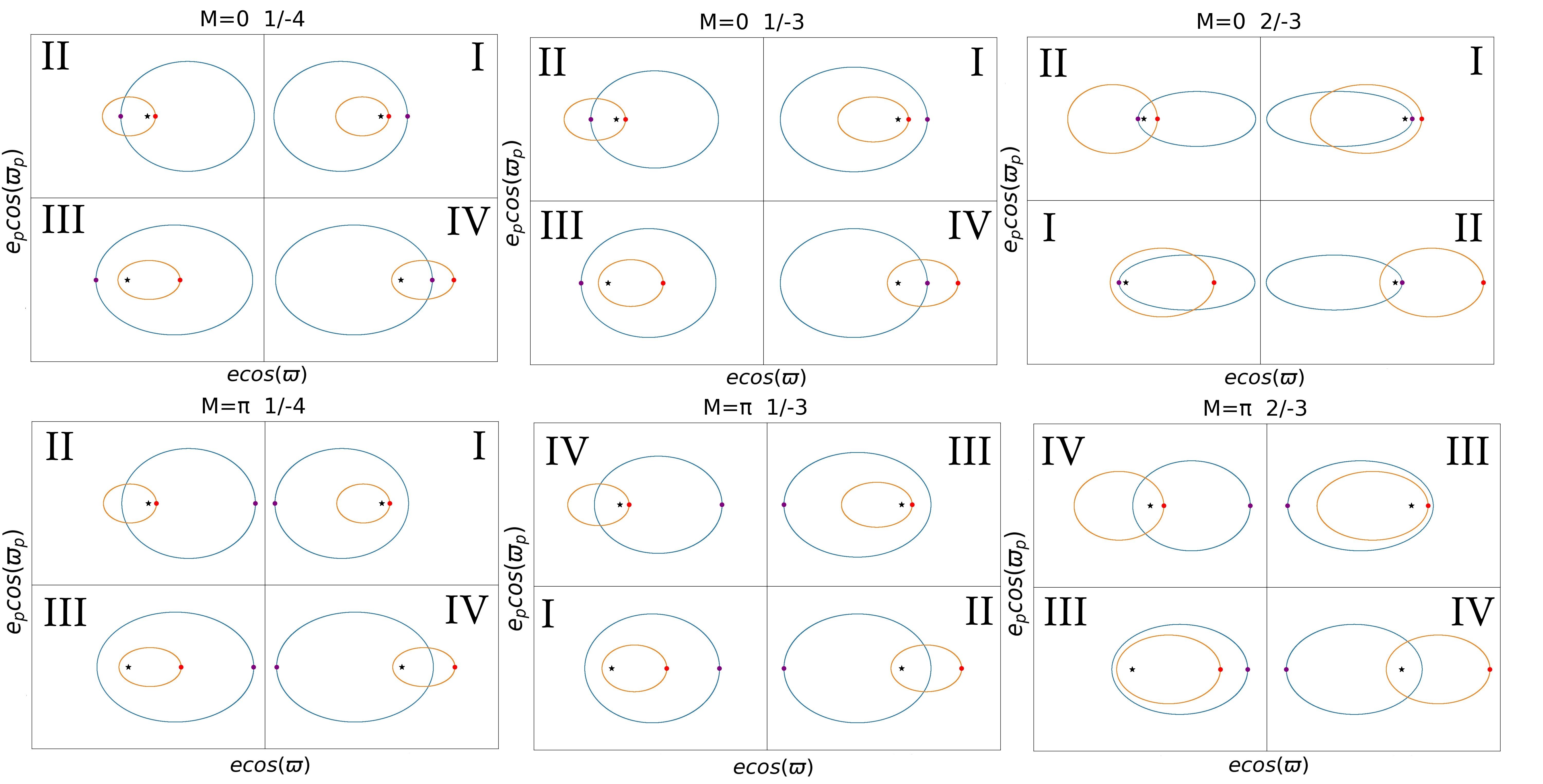}
         \caption{Configurations for the external resonances displayed in 4 quadrants illustrating the orientations of the pericenters and apocenters at initial mean anomalies  $M = 0$ and $M = \pi$. The Roman numerals indicate pairing of the initial configurations which, due to the commensurability between the orbital periods, are equivalent with a time-lag of half a period of the external object (1/-4, 1/-3). For 2/-3 the configurations are equivalent with a time-lag of an entire period.}
         \label{inertialexternal0}
 \end{figure*}

The results of the numerical integrations were arranged in resonant maps computed in a grid of eccentricity of the retrograde body versus  eccentricity of the prograde planet. Using 6400 initial conditions of $(e,e_p)$, we construct a grid of 80x80 initial eccentricities in the range $(0,1)$. For each initial condition, the system was integrated for $2 \times 10^{5} T_p$. To represent all possible configurations, we present our results in maps with four quadrants with a total of 25600 initial conditions, these quadrants correspond to the permutations of aligned / anti-aligned pericenters / apocenters. In $Q_1$, $\varpi_p = \varpi = 0$; in $Q_2$, $\varpi_p = 0, \varpi = \pi$; in $Q_3$, $\varpi_p = \varpi = \pi$, in $Q_4$, $\varpi_p = \pi, \varpi = 0$. In Figure \ref{inertialinner0} and Figure \ref{inertialexternal0} we present respectively the initial configuration for the inner and outer resonances. The initial mean anomaly of the prograde planet has the same value as the longitude of the pericenter $(M_p = \varpi_p)$. The symmetry between the quadrants are identified by the Roman numerals; these indicate which initial conditions are equivalent with a time-lag of half a period of the external object (1/-3, 3/-1, 1/-4, 4/-1) or after a time-lag of an entire period of the external object (2/-3, 3/-2). This equivalence occurs due to the commensurability of the orbital periods but it is not exact due the interaction between the bodies during the time-lag.

\begin{figure}
     \centering
     \begin{subfigure}[b]{0.43\textwidth}
         \centering
         \includegraphics[width=\textwidth]{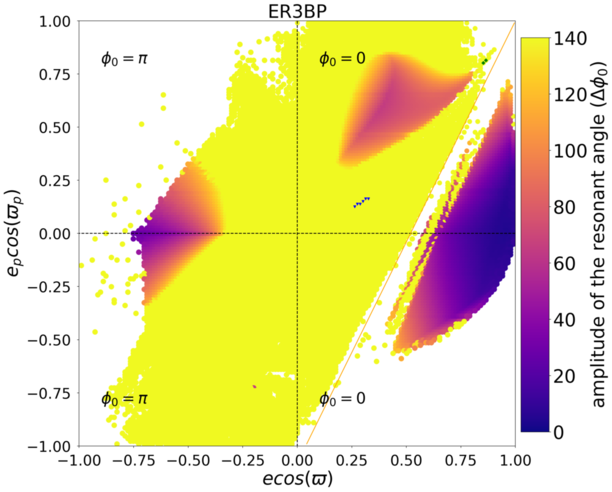}
         \caption{}
         \label{fig:13ER3BP0}
     \end{subfigure}
     \vskip15pt
     \begin{subfigure}[b]{0.43\textwidth}
         \centering
         \includegraphics[width=\textwidth]{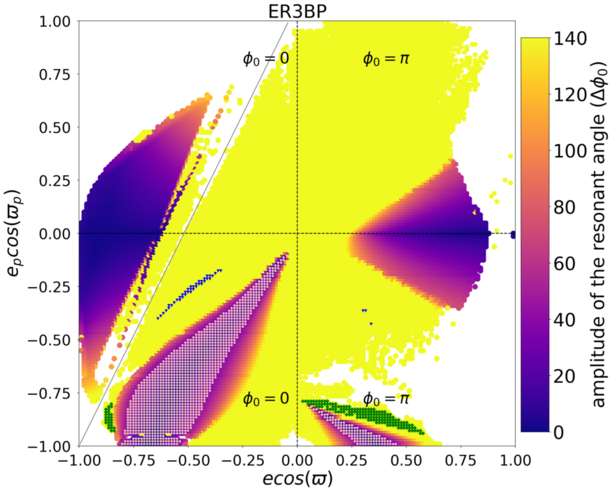} 
         \caption{}
         \label{fig:13ER3BP180}
     \end{subfigure}
     \caption{Resonant maps for the 1/-3 resonance in the elliptic restricted three body problem: (a) $M=0$; (b) $M=\pi$. The amplitude of restricted angle ($\phi_{0}$) is represented by the color bar and the overlaying and the overlaying white symbols indicate the fixed point family where all resonant angles librate around a center. The colored symbols indicate libration of a single resonant angle, $\phi_{2}$ (green) and $\phi_{3}$ (blue). The orange and gray lines indicate, respectively, collision at time zero or after half a period of the external object.}
     \label{fig:131}
     \end{figure}

\begin{figure}
     \centering
     \begin{subfigure}[b]{0.43\textwidth}
         \centering
         \includegraphics[width=\textwidth]{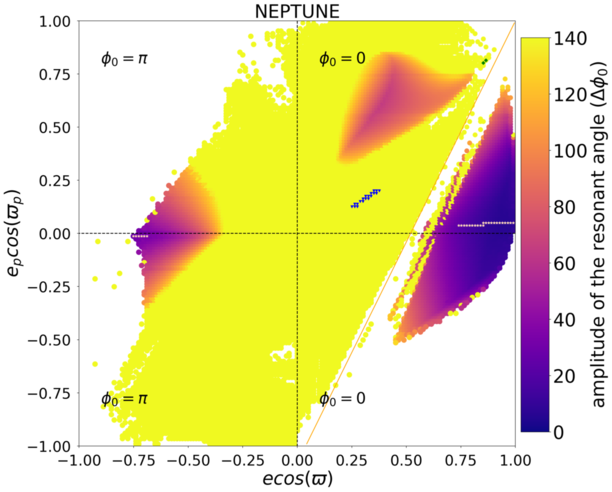} 
         \caption{}
         \label{fig:13NEP0}
     \end{subfigure}
     \vskip15pt
     \begin{subfigure}[b]{0.43\textwidth}
         \centering
         \includegraphics[width=\textwidth]{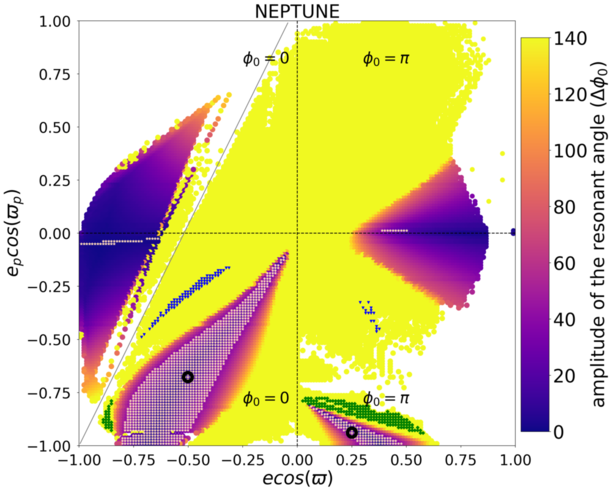} 
         \caption{}
         \label{fig:13NEP180}
     \end{subfigure}
     \caption{Resonant maps for the 1/-3 resonance in the planetary problem when the 2nd planet has Neptune's mass: (a) $M=0$; (b) $M=\pi$. The amplitude of restricted angle ($\phi_{0}$) is represented by the color bar and the overlaying white symbols indicate the fixed point family where all resonant angles librate around a center. The colored symbols indicate libration of a single resonant angle, $\phi_{2}$ (green) and $\phi_{3}$ (blue). The orange and gray lines indicate, respectively, collision at time zero or after half a period of the external object.}     
     \label{fig:132}
     \end{figure}

\begin{figure}
\centering
    \begin{subfigure}[b]{0.43\textwidth}
    \includegraphics[width=\textwidth]{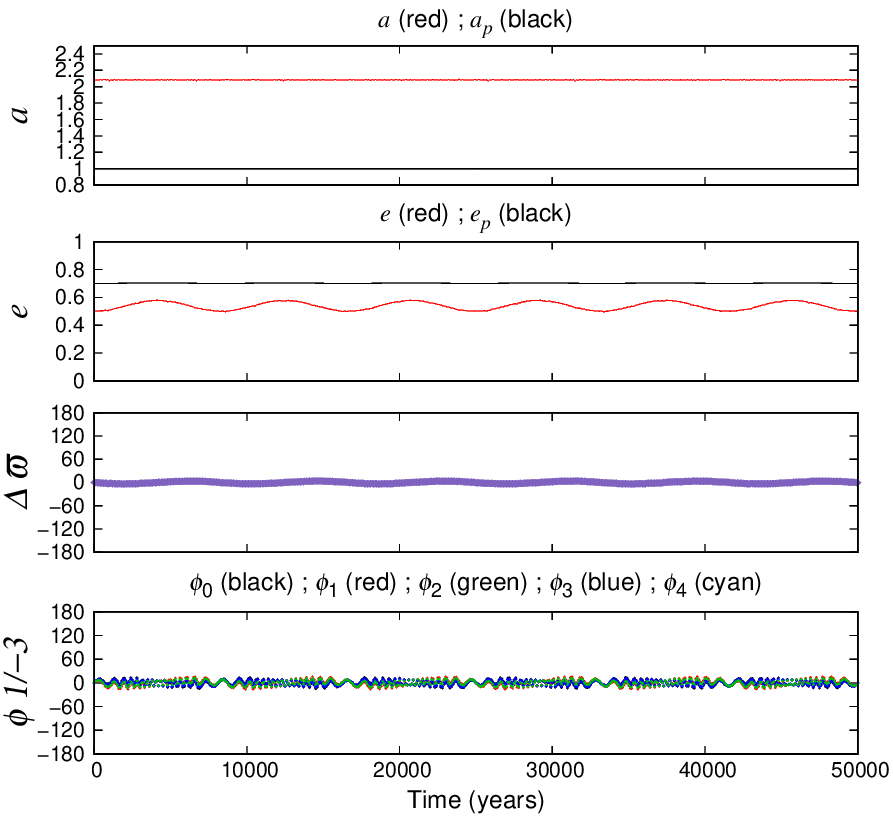}
    \caption{}
    \label{fig:13nepcia}
    \end{subfigure}
    \vskip15pt
    \begin{subfigure}[b]{0.43\textwidth}
      \includegraphics[width=\textwidth]{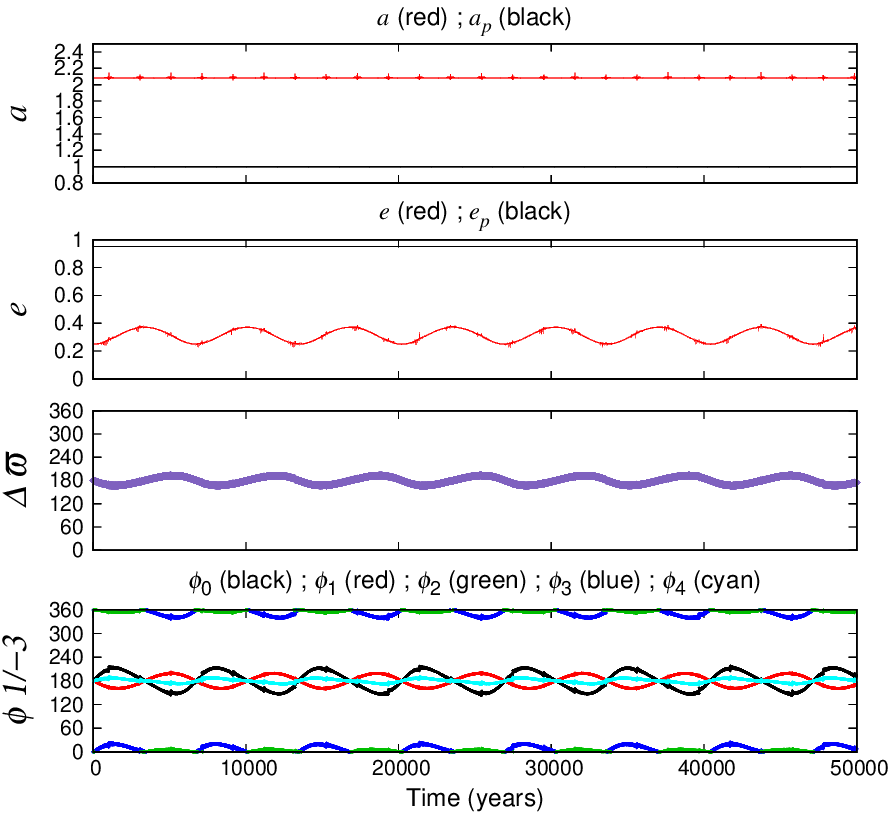}
    \caption{}
    \label{fig:13nepcib}
    \end{subfigure}
    \caption{Orbital evolution of the initial conditions circled in Figure \ref{fig:132}. In (a) the initial condition is $e = 0.5$, $e_p = 0.7$. In (b) the initial condition is $e = 0.25$, $e_p = 0.95$. The 1st, 2nd, 3rd and 4th panels show, respectively, the third body's semi-major axis, its eccentricity, the difference $\Delta \varpi$ between the longitudes of pericenter, and the resonant angles $\phi_0$, $\phi_1$, $\phi_2$, $\phi_3$ and $\phi_4$.}
\label{fig:13nepci}
\end{figure}

\begin{figure}
     \centering
     \begin{subfigure}[b]{0.43\textwidth}
         \centering
         \includegraphics[width=\textwidth]{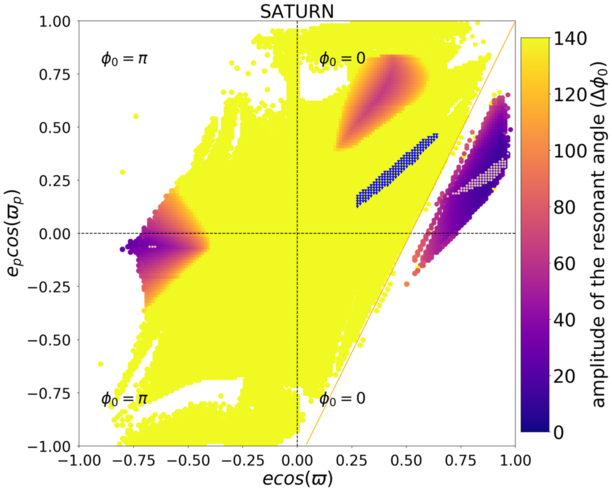}
         \caption{}
         \label{fig:13SAT0}
     \end{subfigure}
     \vskip15pt
     \begin{subfigure}[b]{0.43\textwidth}
         \centering
         \includegraphics[width=\textwidth]{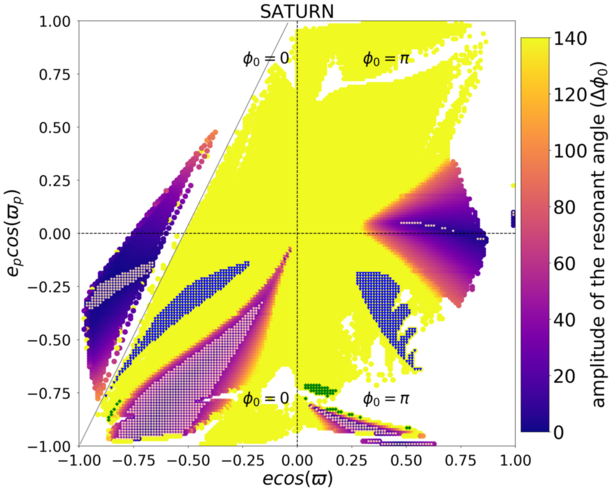}
         \caption{}
         \label{fig:13SAT180}
     \end{subfigure}
     \caption{Resonant maps for the 1/-3 resonance in the planetary problem when the 2nd planet has Saturn's mass: (a) $M=0$; (b) $M=\pi$. The amplitude of restricted angle ($\phi_{0}$) is represented by the color bar and the overlaying white symbols indicate the fixed point family where all resonant angles librate around a center. The colored symbols indicate libration of a single resonant angle, $\phi_{2}$ (green) and $\phi_{3}$ (blue). The orange and gray lines indicate, respectively, collision at time zero or after half a period of the external object.}
     \label{fig:133}
     \end{figure}

\begin{figure}
     \centering
     \begin{subfigure}[b]{0.43\textwidth}
         \centering
         \includegraphics[width=\textwidth]{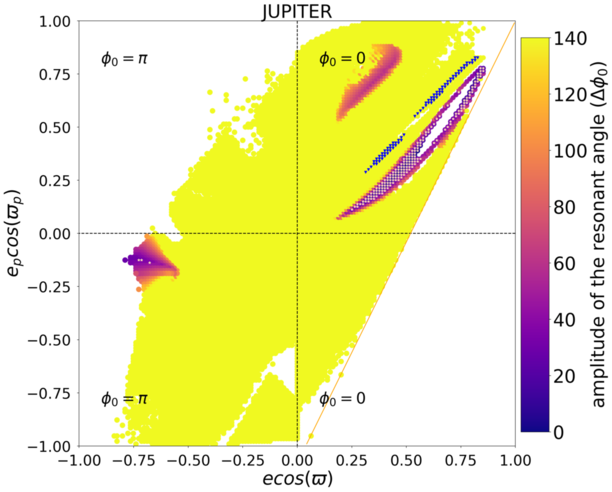} 
         \caption{}
         \label{fig:13JUP0}
     \end{subfigure}
     \vskip15pt
     \begin{subfigure}[b]{0.43\textwidth}
         \centering
         \includegraphics[width=\textwidth]{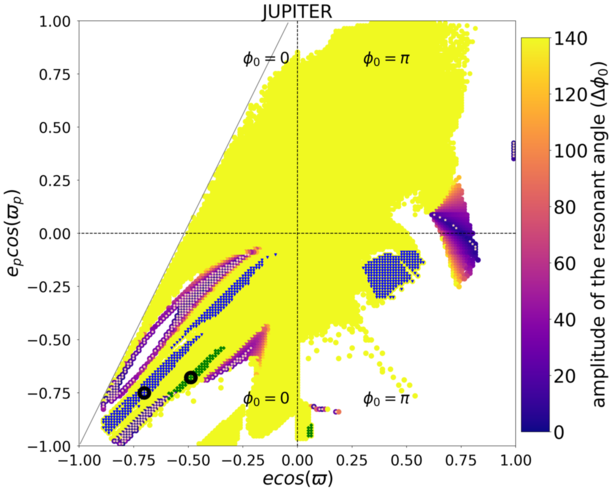}
         \caption{}
         \label{fig:13JUP180}
     \end{subfigure}
     \caption{Resonant maps for the 1/-3 resonance region considering the third body with Jupiter's mass: (a) $M=0$; (b) $M=\pi$. The amplitude of restricted angle ($\phi_{0}$) is represented by the color bar and the overlaying white symbols indicate the fixed point family where all resonant angles librate around a center. The colored symbols indicate libration of a single resonant angle, $\phi_{2}$ (green) and $\phi_{3}$ (blue). The orange and gray lines indicate, respectively, collision at time zero or after half a period of the external object.}
     \label{fig:134}
     \end{figure}

\begin{figure}
\centering
    \begin{subfigure}[b]{0.43\textwidth}
    \includegraphics[width=\textwidth]{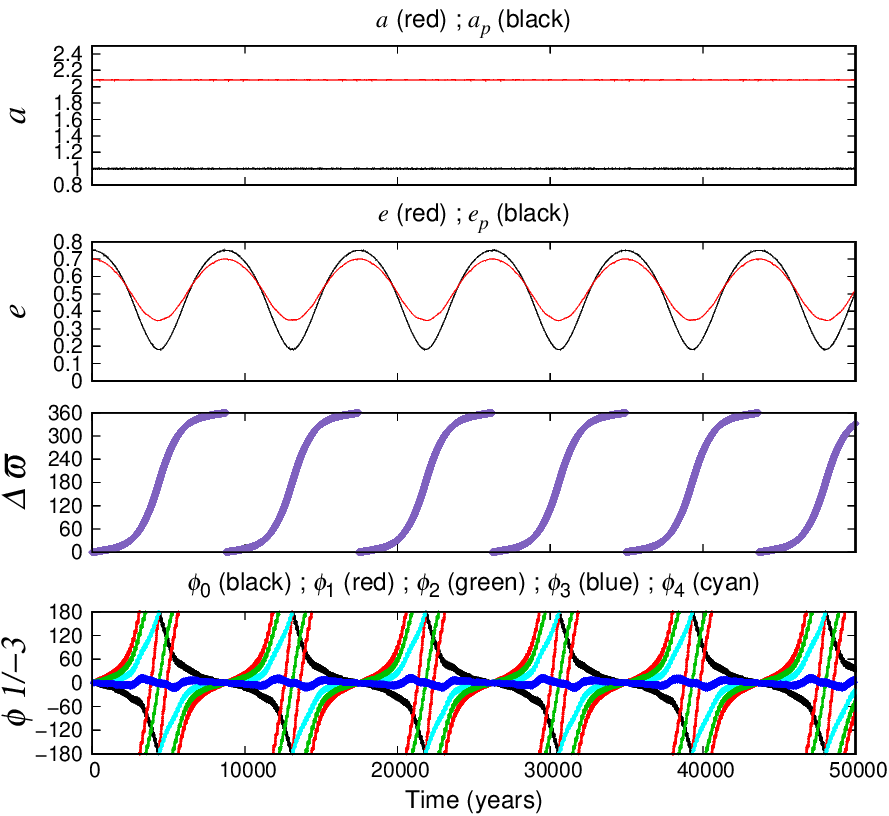}
    \caption{}
    \label{fig:13jupcia}
    \end{subfigure}
    \vskip15pt
    \begin{subfigure}[b]{0.43\textwidth}
      \includegraphics[width=\textwidth]{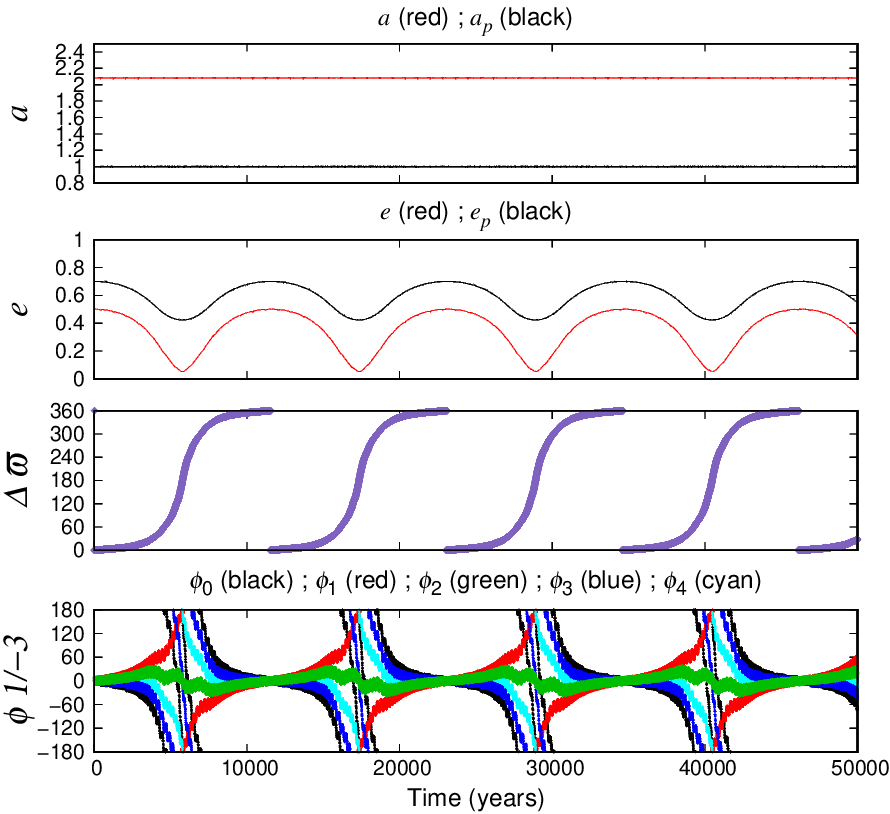}
    \caption{}
    \label{fig:13jupcib}
    \end{subfigure}
    \caption{Orbital evolution of the initial conditions circled in Figure \ref{fig:134}. In (a) the initial condition is $e = 0.7$, $e_p = 0.75$ ($\phi_3$ family). In (b) the initial condition is  $e = 0.5 $, $e_p= 0.7$ ($\phi_2$ family). The 1st, 2nd, 3rd and 4th panels show, respectively, the third body's semi-major axis, its eccentricity, the difference $\Delta \varpi$ between the longitudes of pericenter, and the resonant angles $\phi_0$, $\phi_1$, $\phi_2$, $\phi_3$ and $\phi_4$.}
\label{fig:13jupci}
\end{figure}

\subsection{1/-3 Resonance (Figs. \ref{fig:131}-\ref{fig:13jupci})} 

The resonant angles analyzed were:

\begin{equation}
    \phi_{0} = -3\lambda - \lambda_p + 4\varpi \,\,\, (color \, bar)
\end{equation}
\begin{equation}
    \phi_1 = -3\lambda - \lambda_p + 4\varpi_p   \,\,\, (red)  
\end{equation}
\begin{equation}
   \phi_2 = -3\lambda - \lambda_p + 3\varpi_p + \varpi   \,\,\, (green)
\end{equation}
\begin{equation}
    \phi_3 = -3\lambda - \lambda_p + \varpi_p + 3\varpi   \,\,\, (blue)
\end{equation}
\begin{equation}
    \phi_4 = -3\lambda - \lambda_p + 2\varpi_p + 2\varpi   \,\,\, (purple)
\end{equation}

In Figure \ref{fig:131} for $M = 0$ (a) and $M=\pi$ (b) we present the results for the ER3BP. The amplitude of the resonant angle $\phi_{0}$ is represented by the color bar, where dark purple/blue indicates the resonance center. The green/blue regions indicate the libration of $\phi_2$ and $\phi_3$ respectively. The fixed point families where $\phi_0$ and $\Delta \varpi$ (as well as as all the other resonant angles) librate around 0 or $\pi$ are represented by overlaying white symbols in the darker region. For $M = 0$ (a)  there are no fixed point families but there is libration of $\phi_0$  in all quadrants. In $Q_1$, there are two small colored regions, the color of these indicate single libration of $\phi_2$ (green) and $\phi_3$ (blue). For $M = \pi$ (b), a large fixed point family appears in $Q_3$. Similarly to $Q_1$ for $M = 0$ (a), there are two green/blue regions in $Q_3$ for $M = \pi$. In $Q_4$ there is another fixed family, where $\Delta \varpi$ and three of the resonant angles librate around $\pi$, the other angles librate around $0$. When moving away from the center of the family there is increase of semi-amplitude of the resonant angles and $\Delta \varpi$. The increase in the semi-amplitude of $\phi_2$ libration is smaller than the other angles. The fixed point family is surrounded by a $\phi_2$ single libration region, represented in green. In $Q_4$ there are some initial conditions with libration of $\phi_3$.

The stability maps for the planetary problem when the 2nd planet has Neptune's mass are presented in Figure \ref{fig:132}. In general, the stability regions are similar to the ER3BP maps, however, in the Neptune maps there are fixed point families near $e_p = 0$ which appear in $Q_1$ and $Q_3$ for both values of the mean anomaly. The green/blue regions observed in both panels are larger than in the ER3BP. In Figure \ref{fig:13nepcia} and \ref{fig:13nepcib}, respectively, we show the orbital evolution of the initial conditions marked with black circles in $Q_3$ and $Q_4$ of the Figure \ref{fig:132} (b). For the $Q_3$, a fixed point family with libration around $0$ is presented. In relation to $Q_4$, we present an initial condition of the fixed point family with libration in both centers and $\Delta \varpi$ librates around $\pi$.   

The stability maps for the planetary problem when the 2nd planet has Saturn's mass are presented in Figure \ref{fig:133}. In Figure \ref{fig:133} (a) ($M=0$) the main difference from the case when the 2nd planet has Neptune's mass is that the fixed point region present in $Q_1$ occurs at larger $e_p$. The $\phi_2$ family also disappears for this value of the mass. In Figure \ref{fig:133} (b) ($M=\pi$) we see that the a small stable region for high values of $e$ appears in $Q_1$. In $Q_3$, the fixed point family at large $e$ observed for $e_p \approx 0$ in the Neptune case, is now near $e_p = 0.25$. The fixed point family and the $\phi_2$ family present in $Q_4$ are partially destroyed which indicates that these region are likely to disappear for higher masses. A few initial conditions of fixed points appear near $e_p = 0$, furthermore, a large blue region with libration of $\phi_3$ around $0$ also appears.

The stability maps for the planetary problem when the 2nd planet has Jupiter's mass are presented in Figure \ref{fig:134}. In $Q_1$ for $M = 0$ (a), there is a large fixed point family. However, this family is destroyed for certain values of eccentricity corresponding to the white region in the center of the fixed point family which is vertically unstable. The blue region where libration of $\phi_3$ occurs is divided in two. In $Q_3$ for $M = \pi$ (b), the fixed point family and blue region are similar to the ones described for $M = 0$. In addition, there are two other fixed point families in this quadrant.
The fixed point family with high $e_p$ present in $Q_4$ for the other mass values is almost all destroyed, while the $\phi_3$ libration region reduces in size and the fixed point family with $e_p \approx 0$, observed in Saturn case, is displaced to slightly larger $e_p$. In Figure \ref{fig:13jupci}, we show the orbital evolution of two initial conditions marked in $Q_3$ for $M = \pi$, one of them is maintained by the single libration of $\phi_3$ and the other by the single libration of $\phi_2$. In both of this families there is circulation of $\Delta \varpi$.

\begin{figure}
     \centering
     \begin{subfigure}[b]{0.43\textwidth}
         \centering
         \includegraphics[width=\textwidth]{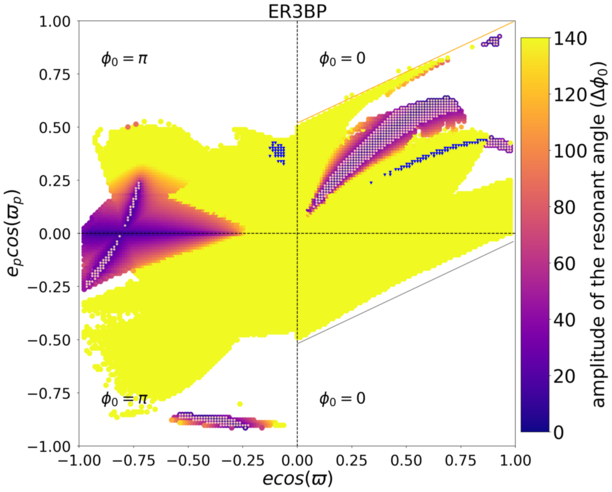}
         \caption{}
         \label{fig:31ER3BP0}
     \end{subfigure}
     \vskip15pt
     \begin{subfigure}[b]{0.43\textwidth}
         \centering
         \includegraphics[width=\textwidth]{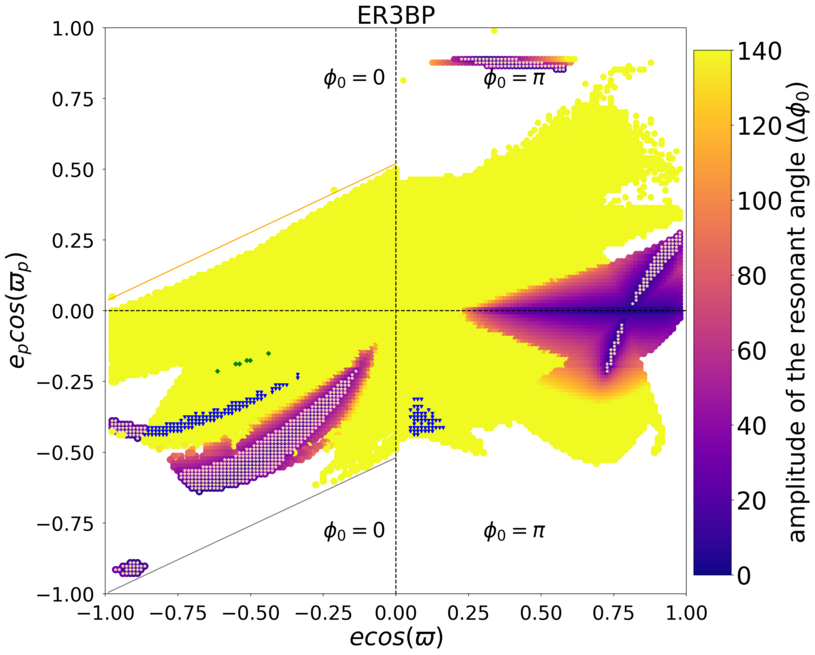}
         \caption{}
         \label{fig:31ER3BP180}
     \end{subfigure}
     \caption{Resonant maps for the 3/-1 resonance in the elliptic restricted three body problem: (a) $M=0$; (b) $M=\pi$. The amplitude of restricted angle ($\phi_{0}$) is represented by the color bar and the overlaying white symbols indicate the fixed point family where all resonant angles librate around a center. The colored symbols indicate libration of a single resonant angle, $\phi_{2}$ (green) and $\phi_{3}$ (blue). The orange and gray lines indicate, respectively, collision at time zero or after half a period of the external object.} 
     \label{fig:311}
     \end{figure}

\begin{figure}
     \centering
     \begin{subfigure}[b]{0.43\textwidth}
         \centering
         \includegraphics[width=\textwidth]{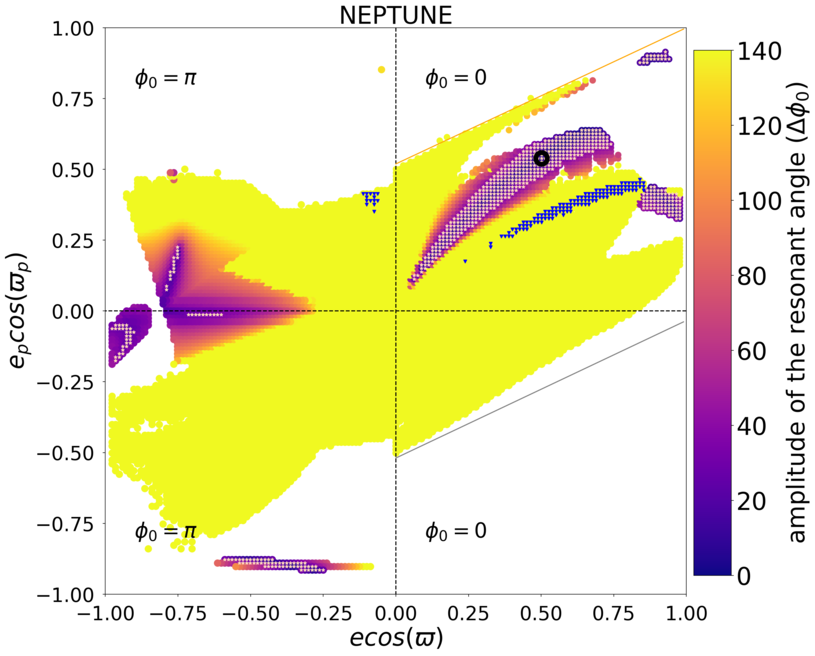}
         \caption{}
         \label{fig:31NEP0}
     \end{subfigure}
     \vskip15pt
     \begin{subfigure}[b]{0.43\textwidth}
         \centering
         \includegraphics[width=\textwidth]{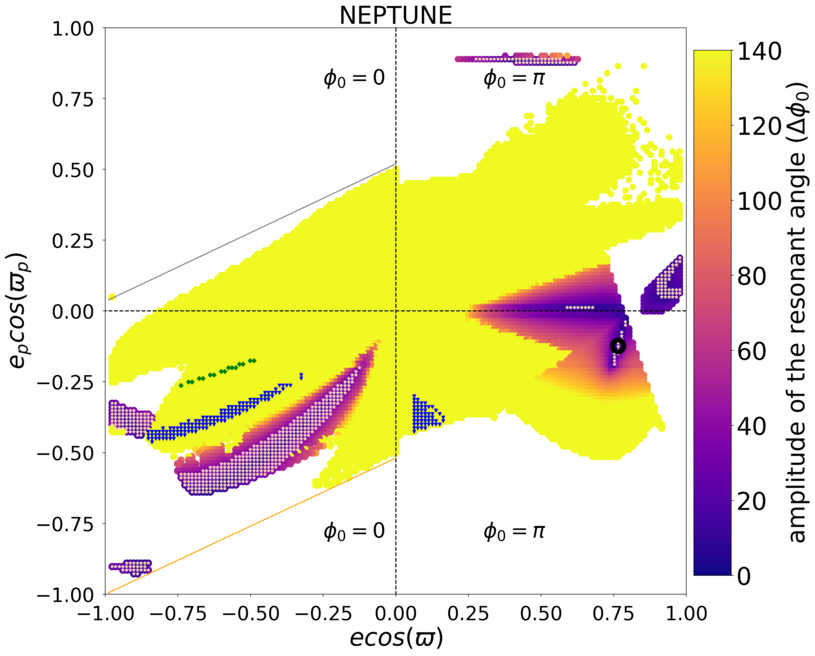} 
         \caption{}
         \label{fig:31NEP180}
     \end{subfigure}
     \caption{Resonant maps for the 3/-1 resonance in the planetary problem when the 2nd planet has Neptune's mass: (a) $M=0$; (b) $M=\pi$. The amplitude of restricted angle ($\phi_{0}$) is represented by the color bar and the overlaying white symbols indicate the fixed point family where all resonant angles librate around a center. The colored symbols indicate libration of a single resonant angle, $\phi_{2}$ (green) and $\phi_{3}$ (blue). The orange and gray lines indicate, respectively, collision at time zero or after half a period of the external object.}
     \label{fig:312}
     \end{figure}

\begin{figure}
\centering
    \begin{subfigure}[b]{0.43\textwidth}
    \includegraphics[width=\textwidth]{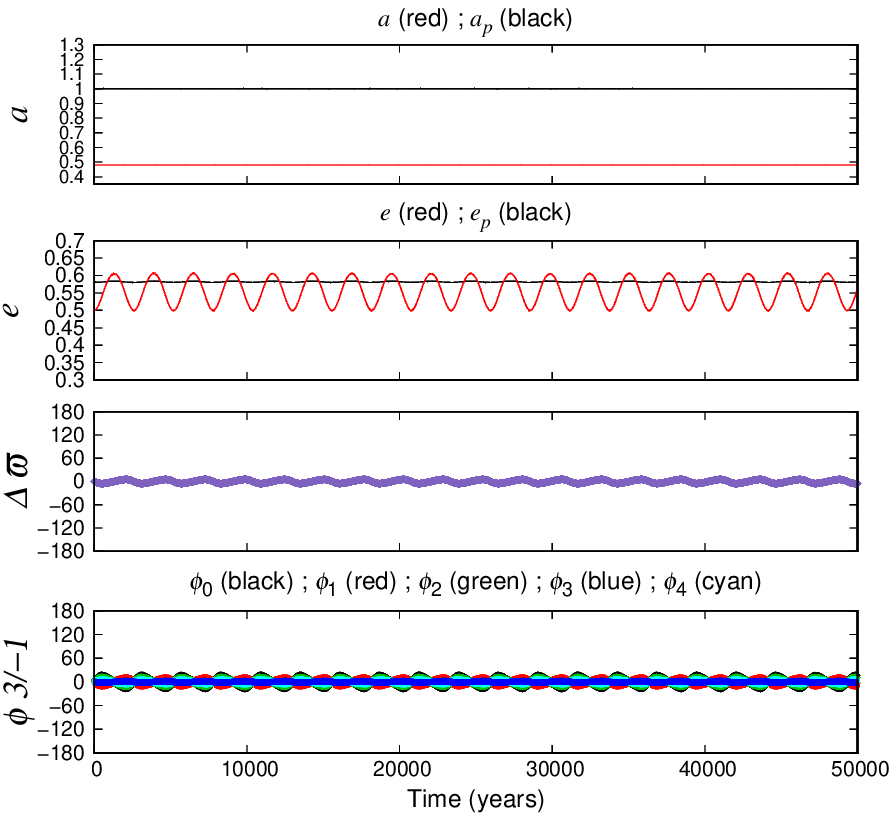}
    \caption{}
    \label{fig:31nepcia}
    \end{subfigure}
    \vskip15pt
    \begin{subfigure}[b]{0.43\textwidth}
      \includegraphics[width=\textwidth]{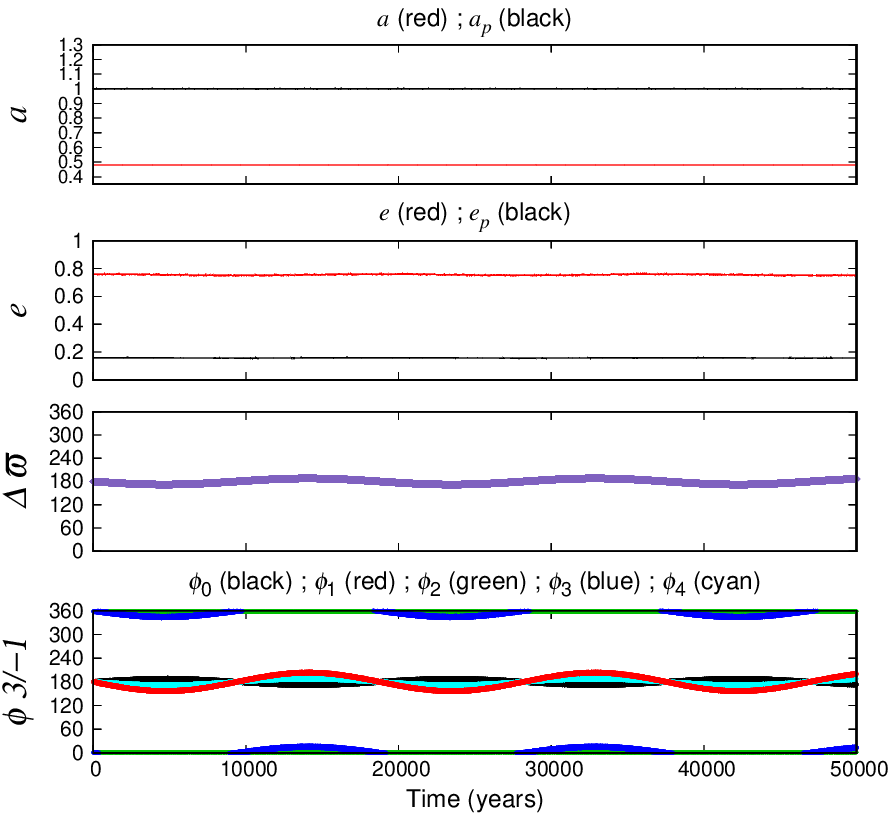}
    \caption{}
    \label{fig:31nepcib}
    \end{subfigure}
    \caption{Orbital evolution of the initial conditions circled in Figure \ref{fig:312}. In (a) the initial condition is $e = 0.5 $, $e_p = 0.58$. In (b) the initial condition is $e = 0.76 $, $e_p = 0.16$. The 1st, 2nd, 3rd and 4th panels show, respectively, the third body's semi-major axis, its eccentricity, the difference $\Delta \varpi$ between the longitudes of pericenter, and the resonant angles $\phi_0$, $\phi_1$, $\phi_2$, $\phi_3$ and $\phi_4$.}
\label{fig:31nepci}
\end{figure}

\begin{figure}
     \centering
     \begin{subfigure}[b]{0.43\textwidth}
         \centering
         \includegraphics[width=\textwidth]{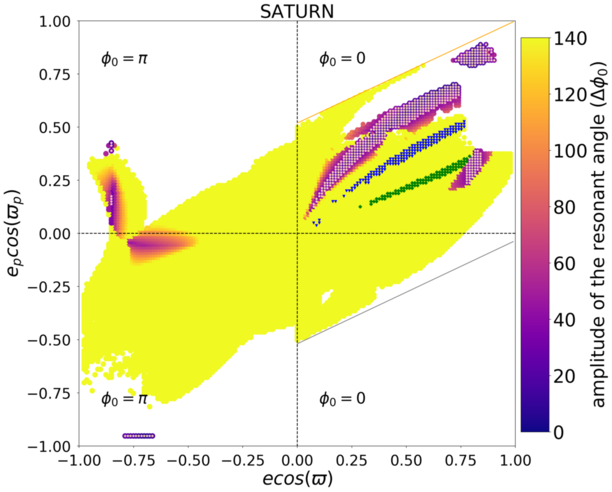}
         \caption{}
         \label{fig:31SAT0}
     \end{subfigure}
     \vskip15pt
     \begin{subfigure}[b]{0.43\textwidth}
         \centering
         \includegraphics[width=\textwidth]{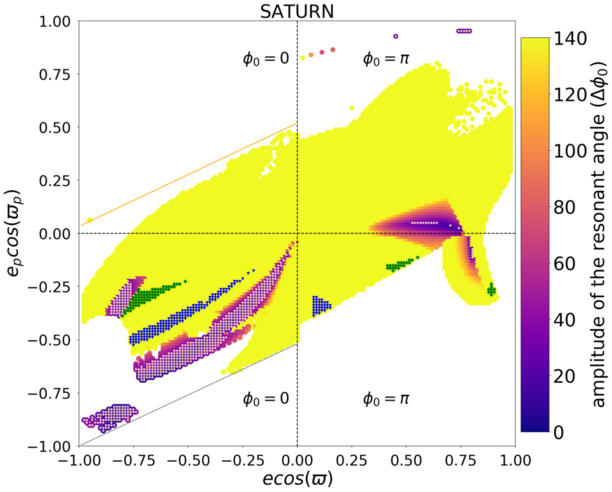}
         \caption{}
         \label{fig:31SAT180}
     \end{subfigure}
     \caption{Resonant maps for the 3/-1 resonance in the planetary problem when the 2nd planet has Saturn's mass: (a) $M=0$; (b) $M=\pi$. The amplitude of restricted angle ($\phi_{0}$) is represented by the color bar and the overlaying white symbols indicate the fixed point family where all resonant angles librate around a center. The colored symbols indicate libration of a single resonant angle, $\phi_{2}$ (green) and $\phi_{3}$ (blue). The orange and gray lines indicate, respectively, collision at time zero or after half a period of the external object.}
     \label{fig:313}
     \end{figure}

\begin{figure}
     \centering
     \begin{subfigure}[b]{0.43\textwidth}
         \centering
         \includegraphics[width=\textwidth]{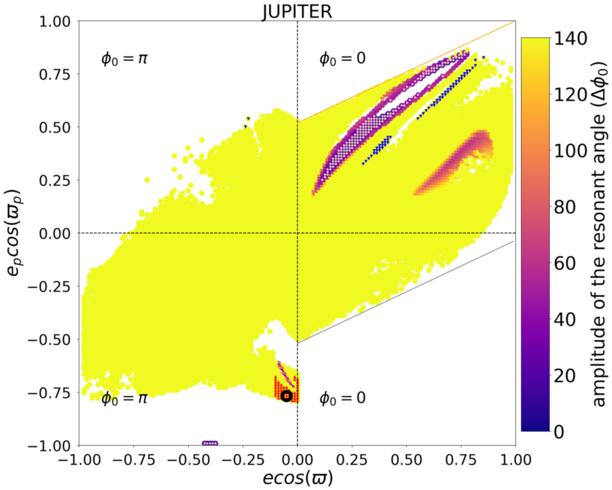}
         \caption{}
         \label{fig:31JUP0}
     \end{subfigure}
     \vskip15pt
     \begin{subfigure}[b]{0.43\textwidth}
         \centering
         \includegraphics[width=\textwidth]{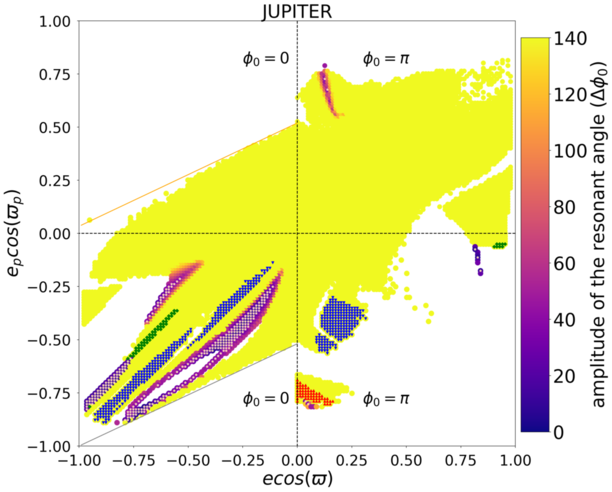}
         \caption{}
         \label{fig:31JUP180}
     \end{subfigure}
     \caption{Resonant maps for the 3/-1 resonance region considering the third body with Jupiter's mass: (a) $M=0$; (b) $M=\pi$. The amplitude of restricted angle ($\phi_{0}$) is represented by the color bar and the overlaying white symbols indicate the fixed point family where all resonant angles librate around a center. The colored symbols indicate libration of a single resonant angle, $\phi_{1}$ (red), $\phi_{2}$ (green) and $\phi_{3}$ (blue). The orange and gray lines indicate, respectively, collision at time zero or after half a period of the external object.}
     \label{fig:314}
     \end{figure}

\begin{figure}
    \centering
    \includegraphics[width=0.43\textwidth]{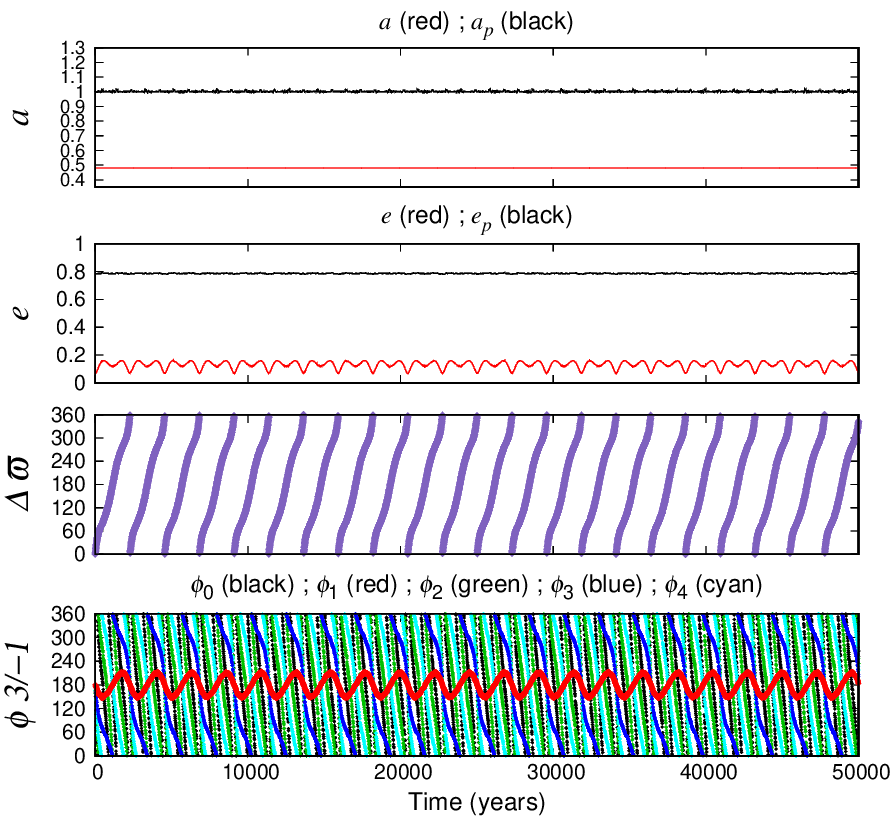}
    \caption{Orbital evolution of the initial condition circled in Figure \ref{fig:314}. The initial condition is $e = 0.07 $, $e_p= 0.78$ and corresponds to an orbit maintained by the libration of $\phi_1$. The 1st, 2nd, 3rd and 4th panels show, respectively, the third body's semi-major axis, its eccentricity, the difference $\Delta \varpi$ between the longitudes of pericenter, and the resonant angles $\phi_0$, $\phi_1$, $\phi_2$, $\phi_3$ and $\phi_4$.}
\label{fig:31jupci}
\end{figure}

\subsection{3/-1 Resonance (Figs. \ref{fig:311}-\ref{fig:31jupci})} 

The resonant angles analyzed were:

\begin{equation}
    \phi_{0} = -\lambda - 3\lambda_p + 4\varpi \,\,\, (\text{color bar})
\end{equation}
\begin{equation}
   \phi_1 = -\lambda - 3\lambda_p + 4\varpi_p   \,\,\, (\text{red})  
\end{equation}
\begin{equation}
  \phi_2 = -\lambda - 3\lambda_p + \varpi_p + 3\varpi   \,\,\, (\text{green})
\end{equation}
\begin{equation}
    \phi_3 = -\lambda - 3\lambda_p + 3\varpi_p + \varpi   \,\,\, (\text{blue})
\end{equation}
\begin{equation}
   \phi_4 = -\lambda - 3\lambda_p + 2\varpi_p + 2\varpi   \,\,\, (\text{cyan})
\end{equation}

In Figure \ref{fig:311}, we present the results for the ER3BP case for $M = 0$ and $M = \pi$. The top panel (a) was obtained for $M = 0$ which implies $\phi_0 = 0$ in $Q_1$, $Q_4$ and $\phi_0 = \pi$ in $Q_2$, $Q_3$. The bottom panel (b) was obtained for $M = \pi$ which implies $\phi_0 = \pi$ in $Q_1$, $Q_4$ and $\phi_0 = 0$ in $Q_2$, $Q_3$. The symmetry between the two maps is evident, for $M = 0$ the fixed point families exist in $Q_1$, $Q_2$ and $Q_3$, while for $M = \pi$ the same fixed point families exist in $Q_1$, $Q_3$ and $Q_4$. We have obtained three fixed point families in $Q_1$ for $M = 0$, moreover there is a $\phi_3$ libration region in this quadrant. The fixed point family of $Q_2$ for $M = 0$ is maintained by the libration of $\phi_2, \phi_3$ around $0$ and $\phi_0, \phi_1, \phi_4$ around $\pi$, in this quadrant there is also a $\phi_3$ libration region. In $Q_3$, there are two fixed point families, both with libration of all angles and of $\Delta \varpi$ around $\pi$. For $M = \pi$, the quadrant $Q_1$ is similar to $Q_3$ for $M = 0$, while $Q_3$ and $Q_4$ have the same fixed point families compared to $Q_1$ and $Q_2$ of the map for $M = 0$. The periodic family observed near to the x-axis in $Q_1$ and $Q_4$ in Figure \ref{fig:311} (b) are in agreement with the fixed point family obtained from bifurcation of the CR3BP by \cite{kotoulas2020planar}. However, the principal fixed point family and the high eccentricity families do not appear in the latter work. Therefore, these new ER3BP families should not occur from bifurcations of the CR3BP. 

The stability maps for the planetary problem when the 2nd planet has Neptune’s mass are presented in Figure \ref{fig:312}. The principal difference from the maps of the ER3BP case  is the separation of the fixed point family observed in $Q_3$ for $M = 0$ and $Q_1$ for $M = \pi$ (Figure \ref{fig:311}) in two families. Another difference is the existence of some initial conditions with libration of $\phi_2$ only in $Q_2$ for $M = 0$. In Figure \ref{fig:31nepci} we show the orbital evolution corresponding to the initial conditions marked by the circles in Figure \ref{fig:312}. The marked circle in $Q_1$ for $M = 0$ corresponds to a fixed point family where all resonant angles and $\Delta \varpi$ librate around 0 (Figure \ref{fig:31nepci} (a)). Figure \ref{fig:31nepci} (b) shows the orbital evolution of the marked circle in $Q_4$ for $M = \pi$ which also corresponds to a fixed point family, however libration of the resonant angles occur around both centers and $\Delta \varpi$ librates around $\pi$.

The stability maps when the 2nd planet has Saturn's mass are presented in Figure \ref{fig:313}. In $Q_1$ of Figure \ref{fig:313} (a) ($M = 0$), there are three fixed point families, a $\phi_3$ libration region, and a $\phi_2$ libration region. The $\phi_2$ libration region were not observed in $M = 0$ map for Neptune's case. The fixed point family reported previously in $Q_2$ for $M = 0$ is nearly destroyed with the mass increase. In $Q_3$ for $M = 0$, only the fixed point family with high $e_p$ survives the increase of mass. In $Q_3$ for $M = \pi$, the fixed point families and the regions of libration of $\phi_2$ and $\phi_3$ survive the increase of mass of the third body. The fixed point family observed in $Q_4$ for $M = \pi$ in Neptune's case is nearly destroyed when the retrograde body has the mass of Saturn. 

The stability maps when the 2nd planet has Jupiter's mass are presented in Figure \ref{fig:314}. In general, there is destruction of  the small fixed point families observed in the Saturn case. The principal fixed point family is still present in $Q_1$ for $M = 0$ and $Q_3$ for $M = \pi$, but for some values of eccentricity this family is vertically unstable. There is a new stable region maintained by the libration of $\phi_1$ around $\pi$, which is present in $Q_3$ for $M = 0$ and $Q_4$ for $M = \pi$. The $\phi_2$ and $\phi_3$ libration regions survive the increase of mass of the third body. In $Q_3$ for $M = 0$, there is a fixed point family for $e \approx 0.4$ and $e_p = 0.99$. In Figure \ref{fig:31jupci} we show the orbital evolution corresponding to the initial condition marked in $Q_3$ for $M = 0$ (Figure \ref{fig:314}) which is maintained by the single libration of $\phi_1$ around $\pi$.

\begin{figure}
     \centering
     \begin{subfigure}[b]{0.43\textwidth}
         \centering
         \includegraphics[width=\textwidth]{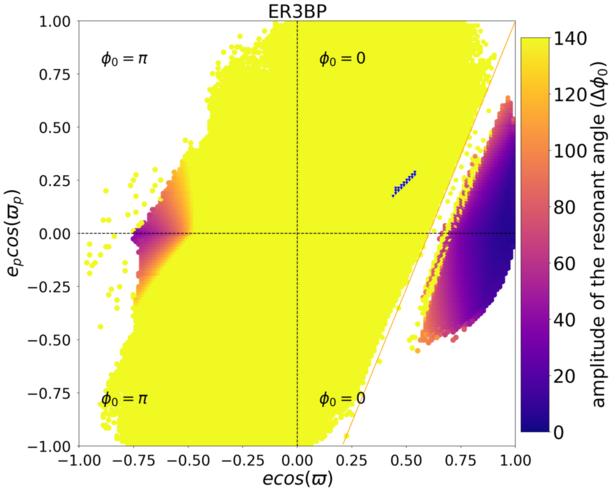}
         \caption{}
         \label{fig:14ER3BP0}
     \end{subfigure}
     \vskip15pt
     \begin{subfigure}[b]{0.43\textwidth}
         \centering
         \includegraphics[width=\textwidth]{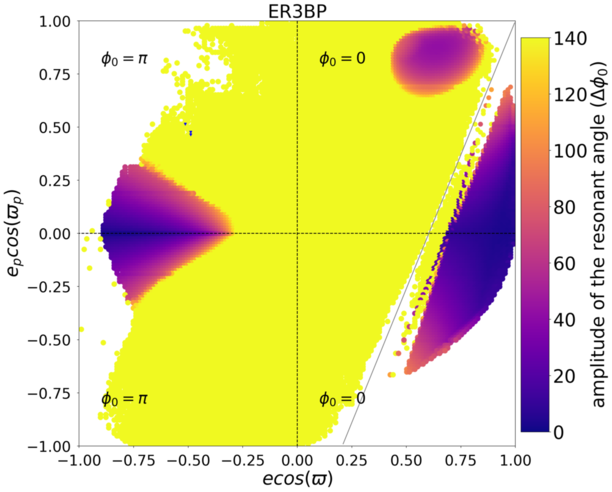}
         \caption{}
         \label{fig:14ER3BP180}
     \end{subfigure}
     \caption{Resonant maps for the 1/-4 resonance in the elliptic restricted three body problem: (a) $M=0$; (b) $M=\pi$. The amplitude of restricted angle ($\phi_{0}$) is represented by the color bar and the overlaying white symbols indicate the fixed point family where all resonant angles librate around a center. The colored symbols indicate libration of a single resonant angle, $\phi_{3}$ (blue). The orange and gray lines indicate, respectively, collision at time zero or after half a period of the external object.} 
     \label{fig:141}
     \end{figure}

\begin{figure}
     \centering
     \begin{subfigure}[b]{0.43\textwidth}
         \centering
         \includegraphics[width=\textwidth]{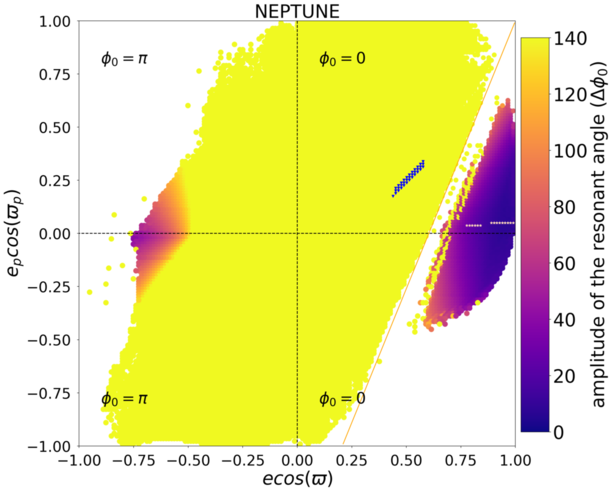}
         \caption{}
         \label{fig:14NEP0}
     \end{subfigure}
     \vskip15pt
     \begin{subfigure}[b]{0.43\textwidth}
         \centering
         \includegraphics[width=\textwidth]{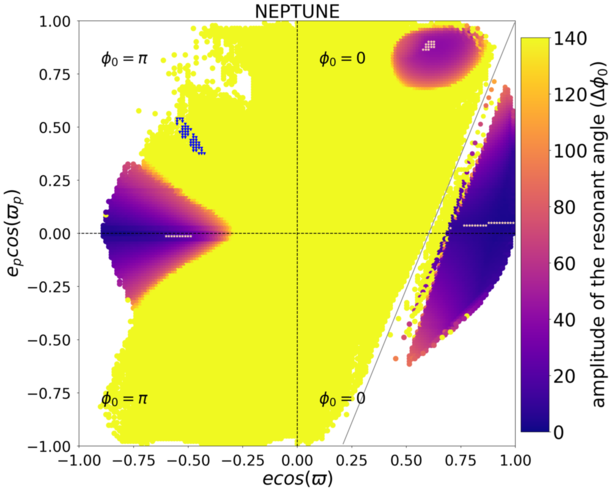}
         \caption{}
         \label{fig:14NEP180}
     \end{subfigure}
     \caption{Resonant maps for the 1/-4 resonance in the planetary problem when the 2nd planet has Neptune's mass: (a) $M=0$; (b) $M=\pi$. The amplitude of restricted angle ($\phi_{0}$) is represented by the color bar and the overlaying white symbols indicate the fixed point family where all resonant angles librate around a center. The colored symbols indicate libration of a single resonant angle, $\phi_{3}$ (blue). The orange and gray lines indicate, respectively, collision at time zero or after half a period of the external object.}
     \label{fig:142}
     \end{figure}

\begin{figure}
     \centering
     \begin{subfigure}[b]{0.43\textwidth}
         \centering
         \includegraphics[width=\textwidth]{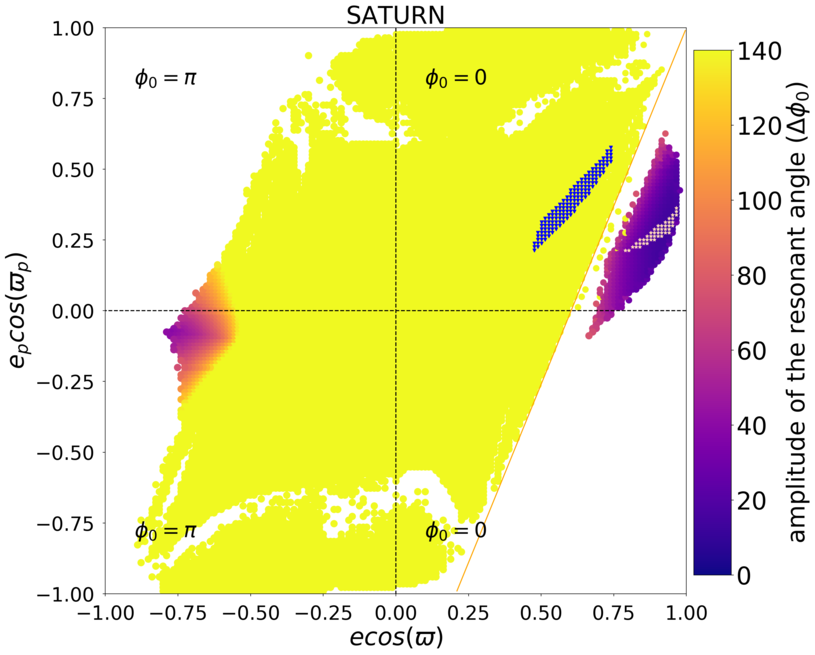}   
         \caption{}
         \label{fig:14SAT0}
     \end{subfigure}
     \vskip15pt
     \begin{subfigure}[b]{0.43\textwidth}
         \centering
         \includegraphics[width=\textwidth]{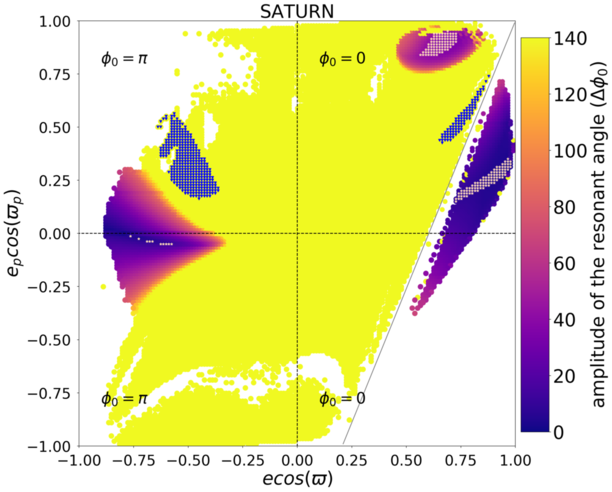}
         \caption{}
         \label{fig:14SAT180}
     \end{subfigure}
     \caption{Resonant maps for the 1/-4 resonance in the planetary problem when the 2nd planet has Saturn's mass: (a) $M=0$; (b) $M=\pi$. The amplitude of restricted angle ($\phi_{0}$) is represented by the color bar and the overlaying white symbols indicate the fixed point family where all resonant angles librate around a center. The colored symbols indicate libration of a single resonant angle, $\phi_{3}$ (blue). The orange and gray lines indicate, respectively, collision at time zero or after half a period of the external object.}
     \label{fig:143}
     \end{figure}

\begin{figure}
     \centering
     \begin{subfigure}[b]{0.43\textwidth}
         \centering
         \includegraphics[width=\textwidth]{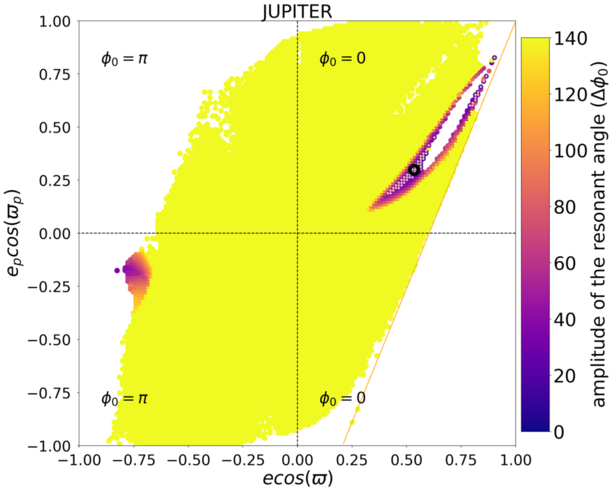}
         \caption{}
         \label{fig:14JUP0}
     \end{subfigure}
     \vskip15pt
     \begin{subfigure}[b]{0.43\textwidth}
         \centering
         \includegraphics[width=\textwidth]{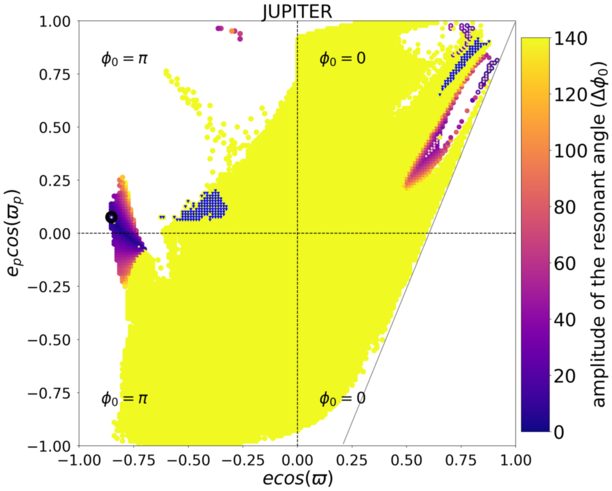}
         \caption{}
         \label{fig:14JUP180}
     \end{subfigure}
     \caption{Resonant maps for the 1/-4 resonance region considering the third body with Jupiter's mass: (a) $M=0$; (b) $M=\pi$. The amplitude of restricted angle ($\phi_{0}$) is represented by the color bar and the overlaying white symbols indicate the fixed point family where all resonant angles librate around a center. The colored symbols indicate libration of a single resonant angle, $\phi_{3}$ (blue). The orange and gray lines indicate, respectively, collision at time zero or after half a period of the external object.}
     \label{fig:144}
     \end{figure}

\begin{figure}
\centering
    \begin{subfigure}[b]{0.43\textwidth}
    \includegraphics[width=\textwidth]{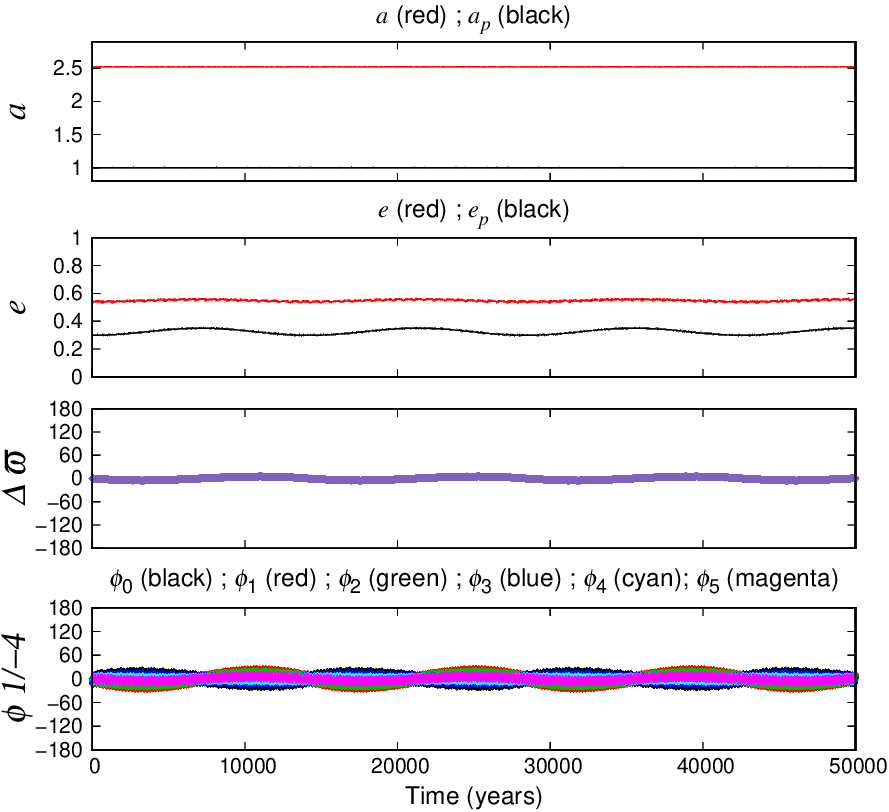}
    \caption{}
    \label{fig:14jupcia}
    \end{subfigure}
    \vskip15pt
    \begin{subfigure}[b]{0.43\textwidth}
      \includegraphics[width=\textwidth]{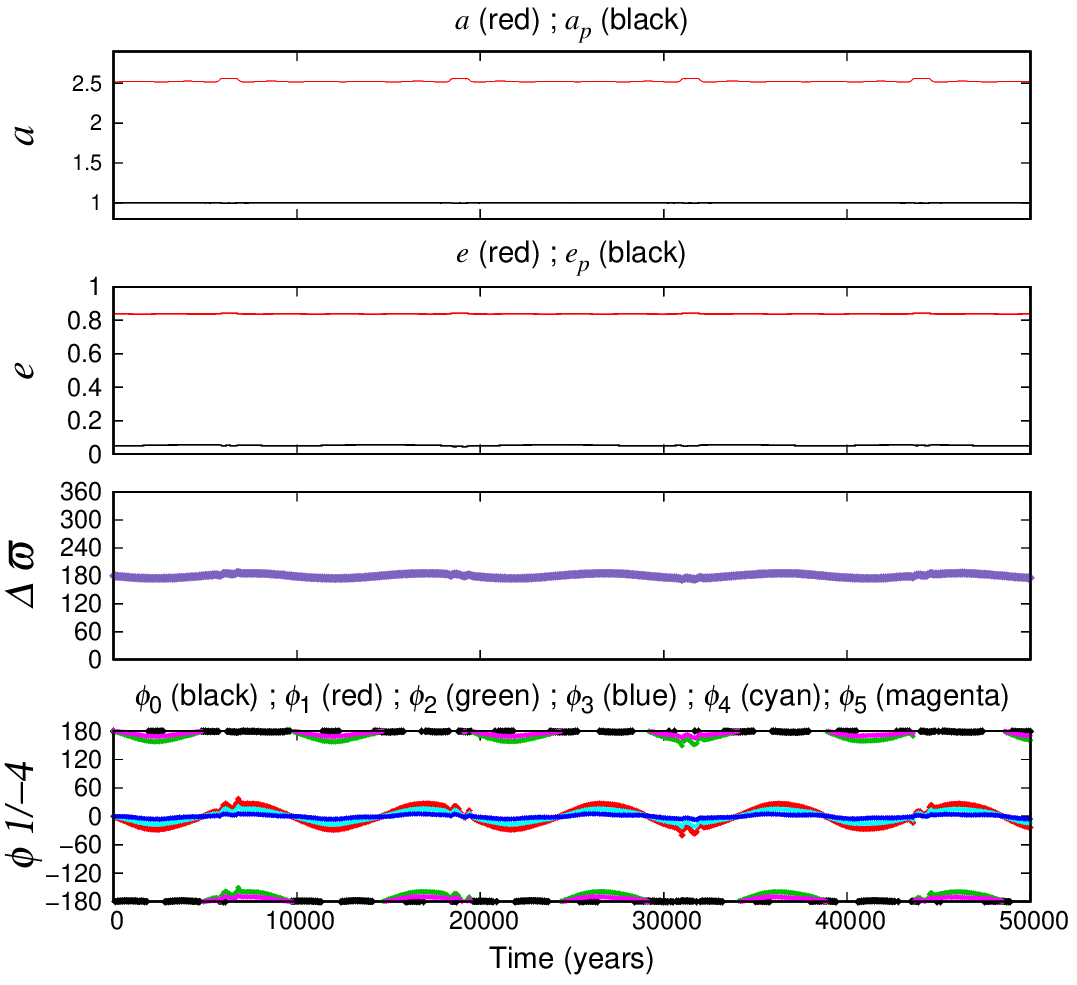}
    \caption{}
    \label{fig:14jupcib}
    \end{subfigure}
    \caption{Orbital evolution of the initial conditions circled in Figure \ref{fig:144}. In (a) the initial condition is $e = 0.54 $, $e_p = 0.3$. In (b) the initial condition is $e = 0.84 $, $e_p = 0.05$. The 1st, 2nd, 3rd and 4th panels show, respectively, the third body's semi-major axis, its eccentricity, the difference $\Delta \varpi$ between the longitudes of pericenter, and the resonant angles $\phi_0$, $\phi_1$, $\phi_2$, $\phi_3$, $\phi_4$ and $\phi_5$.}
\label{fig:14jupci}
\end{figure}

\subsection{1/-4 Resonance (Figs. \ref{fig:141}-\ref{fig:14jupci})} 

The resonant angles analyzed were:

\begin{equation}
    \phi_{0} = -4\lambda - \lambda_p + 5\varpi \,\,\, (\text{color bar})
\end{equation}
\begin{equation}
    \phi_1 = -4\lambda - \lambda_p + 5\varpi_p   \,\,\, (\text{red})   
\end{equation}
\begin{equation}
   \phi_2 = -4\lambda - \lambda_p + 4\varpi_p + \varpi   \,\,\, (\text{green})
\end{equation}
\begin{equation}
    \phi_3 = -4\lambda - \lambda_p + \varpi_p + 4\varpi   \,\,\, (\text{blue})
\end{equation}
\begin{equation}
    \phi_4 = -4\lambda - \lambda_p + 3\varpi_p + 2\varpi   \,\,\, (\text{cyan})
\end{equation}
\begin{equation}
    \phi_5 = -4\lambda - \lambda_p + 2\varpi_p + 3\varpi   \,\,\, (\text{magenta})
\end{equation}

In Figure \ref{fig:141} for $M = 0$ (a) and $M = \pi$ (b) we present the stability maps for the ER3BP. The color bar indicates the amplitude of the resonant angle $\phi_0$ where dark purple/blue indicates the resonance center. For both values of mean anomaly, we did not observe fixed point families in the maps. In $Q_1$ for $M = 0$ and $Q_2$ for $M = \pi$, there is a region where only the resonant angle $\phi_3$ librates around $0$ while the others angles circulate. For $M = \pi$, there is a region of $\phi_0$ libration in the case the prograde planet has high eccentricity ($e_p > 0.75$). Some initial conditions have libration of $\phi_3$ around $0$ in $Q_2$ of Figure \ref{fig:141} (b).

The stability maps for the planetary problem when the 2nd planet has Neptune mass are presented in Figure \ref{fig:142}. As observed for the other resonances, the main difference between the restricted and the Neptune case is related to the appearance of fixed point families near $e_p \approx 0$. This occurs for the 1/-4 resonance, however we also observe the emergence of a fixed point family with $e_p \approx 0.88$ and $e \approx 0.6$ in the Neptune case. A $\phi_3$ libration region around 0 appears in $Q_2$ for $M = \pi$.

The stability maps for the planetary problem when the third body has Saturn mass are presented in Figure \ref{fig:143}. Comparing to the maps with Neptune mass, we can see that the fixed point families with $e_p \approx 0$ in $Q_1$ and $Q_3$ are displaced to higher values of the prograde planet eccentricity. The two fixed point families in $Q_1$ also increase in size. For $M = \pi$, a new region of $\phi_3$ libration appears in $Q_1$, while the other $\phi_3$ libration regions observed in both maps ($Q_1$ for $M = 0$ and $Q_2$ for $M = \pi$) grow in size with increasing mass. In $Q_1$, the fixed point family with $e_p = 0.88$ also grow in size when the retrograde planet have the mass of Saturn. 

In Figure \ref{fig:144}, the stability maps when the second planet has Jupiter mass are presented. In this case, we can observe that in $Q_1$ of both maps ($M = 0$ and $M = \pi$) the most part of the  fixed point regions are destroyed due to vertical instability. For $M = 0$ there is no $\phi_3$ libration region; for $M = \pi$ these regions are still present, one of them at very high $e_p$. Differently from the previous maps, there is a periodic family near the x-axis in $Q_2$ for $M = \pi$. In Figure \ref{fig:14jupci} we show the orbital evolution of two initial conditions marked in Figure \ref{fig:144}. Both initial conditions correspond to fixed points, the first one, represented in Figure \ref{fig:14jupci} (a), exhibits libration of all resonant angles and $\Delta \varpi$ around $0$; the second one, represented in Figure \ref{fig:14jupci} (b), exhibits libration of $\phi_0, \phi_2, \phi_5$ around $0$ and $\phi_1, \phi_3, \phi_4, \Delta \varpi$ around $\pi$.

\begin{figure}
     \centering
     \begin{subfigure}[b]{0.43\textwidth}
         \centering
         \includegraphics[width=\textwidth]{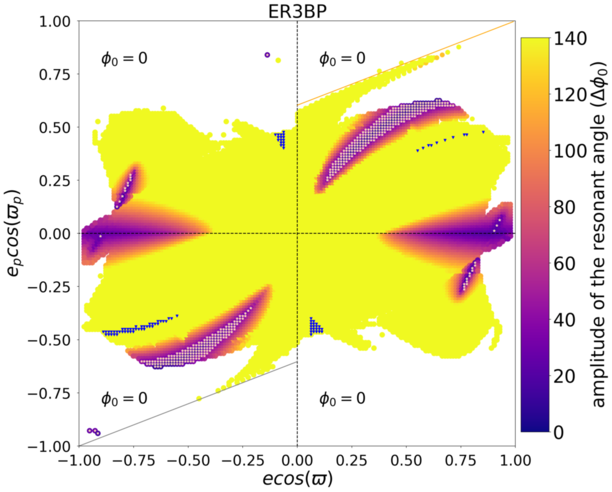}
         \caption{}
         \label{fig:41ER3BP0}
     \end{subfigure}
     \vskip15pt
     \begin{subfigure}[b]{0.43\textwidth}
         \centering
         \includegraphics[width=\textwidth]{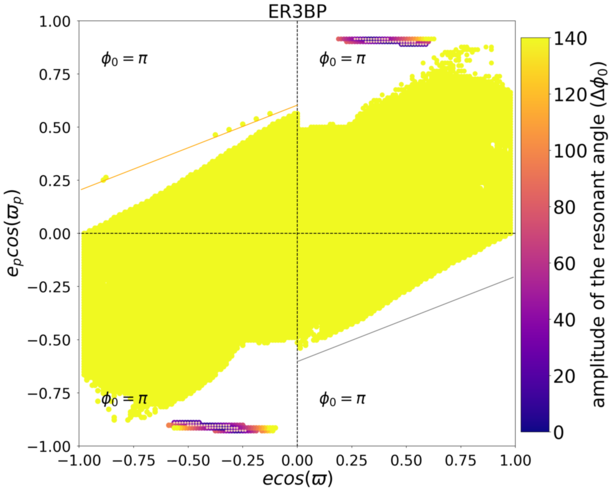}
         \caption{}
         \label{fig:41ER3BP180}
     \end{subfigure}
     \caption{Resonant maps for the 4/-1 resonance in the elliptic restricted three body problem: (a) $M=0$; (b) $M=\pi$. The amplitude of restricted angle ($\phi_{0}$) is represented by the color bar and the overlaying white symbols indicate the fixed point family where all resonant angles librate around a center. The colored symbols indicate libration of a single resonant angle, $\phi_{3}$ (blue). The orange and gray lines indicate, respectively, collision at time zero or after half a period of the external object.}
     \label{fig:411}
     \end{figure}

\begin{figure}
     \centering
     \begin{subfigure}[b]{0.43\textwidth}
         \centering
         \includegraphics[width=\textwidth]{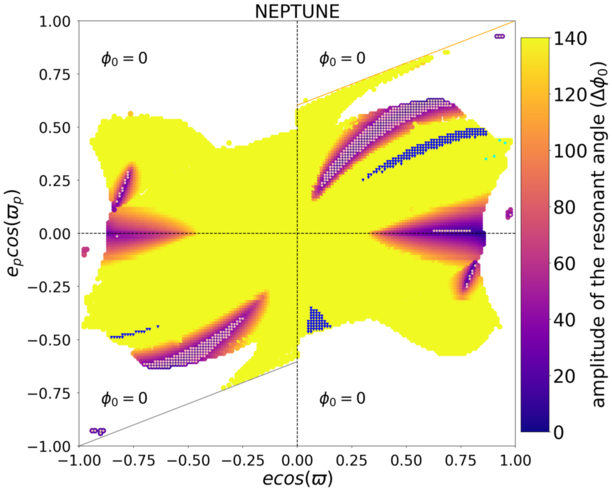}
         \caption{}
         \label{fig:41NEP0}
     \end{subfigure}
     \vskip15pt
     \begin{subfigure}[b]{0.43\textwidth}
         \centering
         \includegraphics[width=\textwidth]{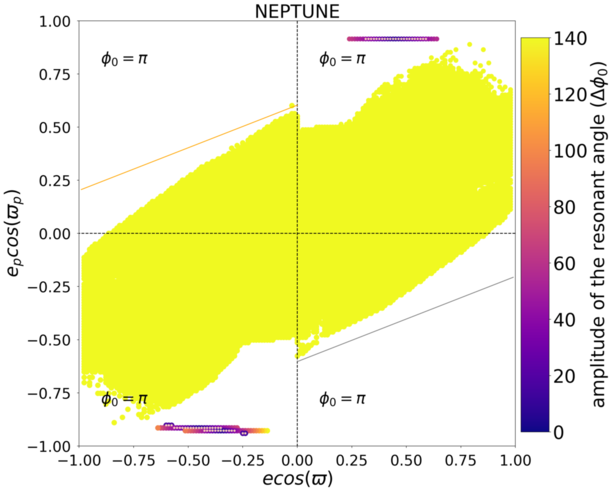}
         \caption{}
         \label{fig:41NEP180}
     \end{subfigure}
     \caption{Resonant maps for the 4/-1 resonance in the planetary problem when the 2nd planet has Neptune's mass: (a) $M=0$; (b) $M=\pi$. The amplitude of restricted angle ($\phi_{0}$) is represented by the color bar and the overlaying white symbols indicate the fixed point family where all resonant angles librate around a center. The colored symbols indicate libration of a single resonant angle, $\phi_{3}$ (blue). The orange and gray lines indicate, respectively, collision at time zero or after half a period of the external object.}
     \label{fig:412}
     \end{figure}

\begin{figure}
     \centering
     \begin{subfigure}[b]{0.43\textwidth}
         \centering
         \includegraphics[width=\textwidth]{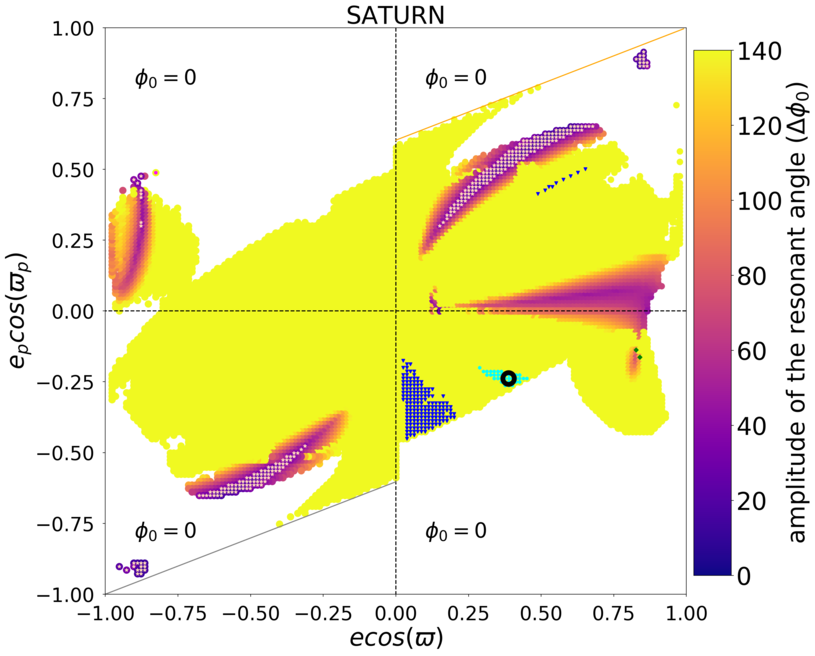}  
         \caption{}
         \label{fig:41SAT0}
     \end{subfigure}
     \vskip15pt
     \begin{subfigure}[b]{0.43\textwidth}
         \centering
         \includegraphics[width=\textwidth]{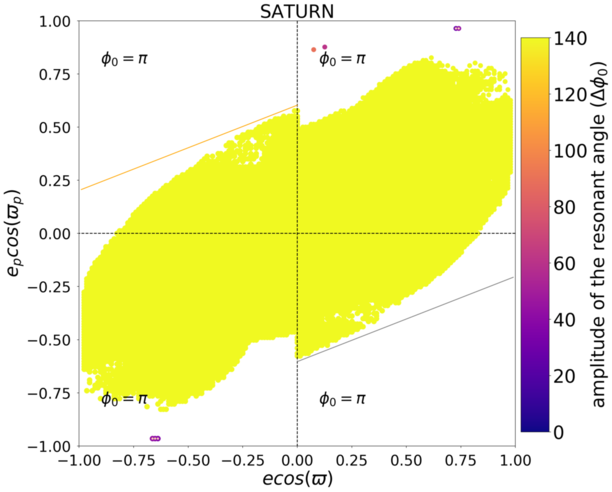}
         \caption{}
         \label{fig:41SAT180}
     \end{subfigure}
     \caption{Resonant maps for the 4/-1 resonance in the planetary problem when the 2nd planet has Saturn's mass: (a) $M=0$; (b) $M=\pi$. The amplitude of restricted angle ($\phi_{0}$) is represented by the color bar and the overlaying white symbols indicate the fixed point family where all resonant angles librate around a center. The colored symbols indicate libration of a single resonant angle, $\phi_{3}$ (blue) and $\phi_4$ (cyan). The orange and gray lines indicate, respectively, collision at time zero or after half a period of the external object.}
     \label{fig:413}
     \end{figure}

\begin{figure}
    \centering
    \includegraphics[width=0.43\textwidth]{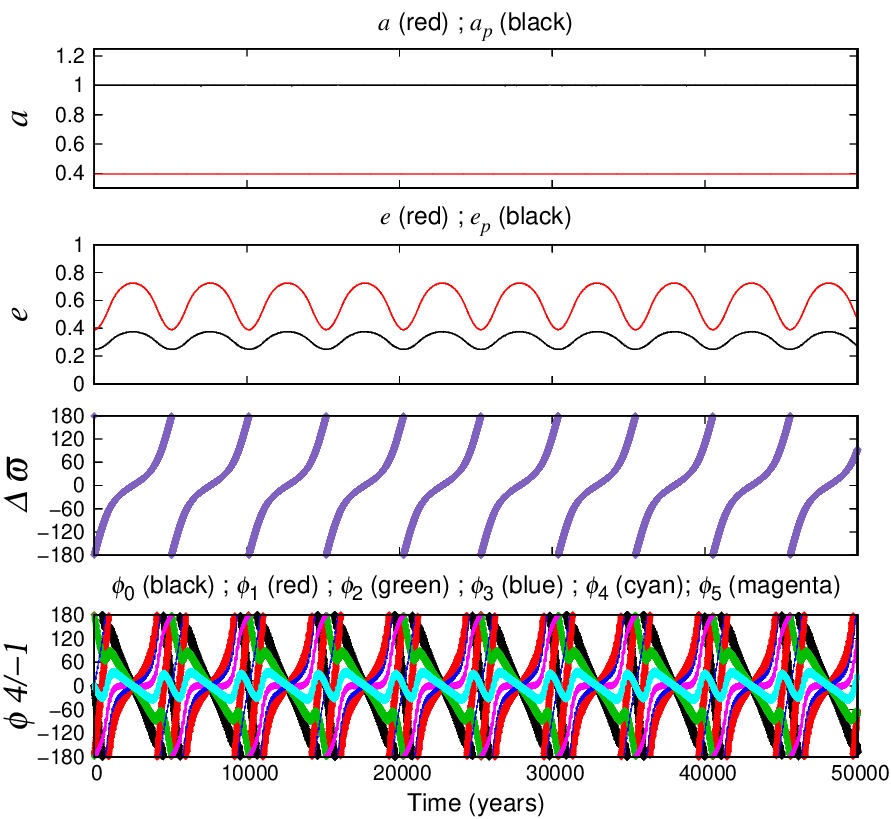}
    \caption{Orbital evolution of the initial condition circled in Figure \ref{fig:413}. The initial condition is $e = 0.39 $, $e_p = 0.25$ and corresponds to an orbit maintained by the libration of $\phi_4$. The 1st, 2nd, 3rd and 4th panels show, respectively, the third body's semi-major axis, its eccentricity, the difference $\Delta \varpi$ between the longitudes of pericenter, and the resonant angles $\phi_0$, $\phi_1$, $\phi_2$, $\phi_3$, $\phi_4$ and $\phi_5$.}
\label{fig:41satci}
\end{figure}

\begin{figure}
     \centering
     \begin{subfigure}[b]{0.43\textwidth}
         \centering
         \includegraphics[width=\textwidth]{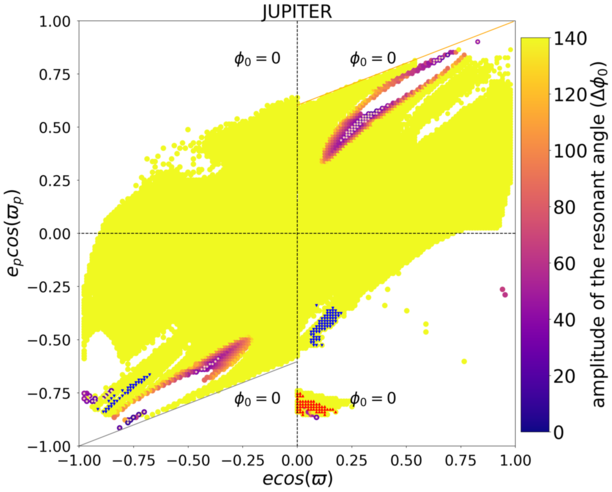}
         \caption{}
         \label{fig:41JUP0}
     \end{subfigure}
     \vskip15pt
     \begin{subfigure}[b]{0.43\textwidth}
         \centering
         \includegraphics[width=\textwidth]{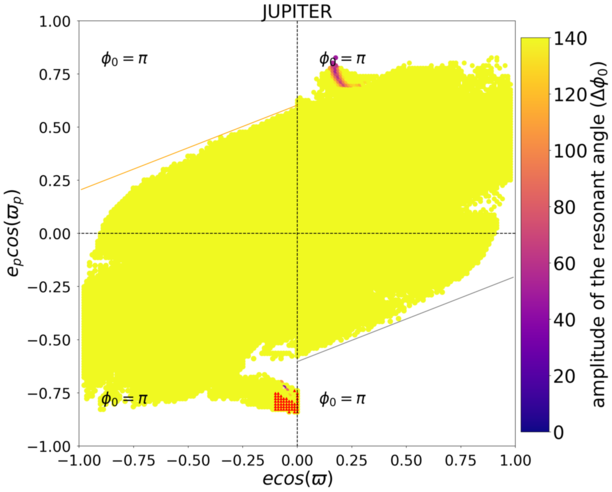}
         \caption{}
         \label{fig:41JUP180}
     \end{subfigure}
     \caption{Resonant maps for the 4/-1 resonance region considering the third body with Jupiter's mass: (a) $M=0$; (b) $M=\pi$. The amplitude of restricted angle ($\phi_{0}$) is represented by the color bar and the overlaying white symbols indicate the fixed point family where all resonant angles librate around a center. The colored symbols indicate libration of a single resonant angle, $\phi_1$ (red) and $\phi_{3}$ (blue). The orange and gray lines indicate, respectively, collision at time zero or after half a period of the external object.}
     \label{fig:414}
     \end{figure}

\subsection{4/-1 Resonance (Figs. \ref{fig:411}-\ref{fig:414})} 

The resonant angles analyzed were:

\begin{equation}
    \phi_{0} = -\lambda - 4\lambda_p + 5\varpi \,\,\, (\text{color bar})
\end{equation}
\begin{equation}
    \phi_1 = -\lambda - 4\lambda_p + 5\varpi_p   \,\,\, (\text{red})  
\end{equation}
\begin{equation}
   \phi_2 = -\lambda - 4\lambda_p + \varpi_p + 4\varpi   \,\,\, (\text{green})
\end{equation}
\begin{equation}
    \phi_3 = -\lambda - 4\lambda_p + 4\varpi_p + \varpi   \,\,\, (\text{blue})
\end{equation}
\begin{equation}
    \phi_4 = -\lambda - 4\lambda_p + 2\varpi_p + 3\varpi   \,\,\, (\text{cyan})
\end{equation}
\begin{equation}
    \phi_5 = -\lambda - 4\lambda_p + 3\varpi_p + 2\varpi   \,\,\, (\text{magenta})
\end{equation}

In Figure \ref{fig:411} we present the maps for the ER3BP. The top panel (a) correspond to $M = 0$ which implies $\phi_0 = 0$ in all quadrants, while the bottom panel (b) correspond to $M = \pi$ implying $\phi_0 = \pi$ in all quadrants. As we can see on the maps, the symmetry in this resonance occurs across the x-axis. For $M = 0$ map, there are periodic families and $\phi_3$ libration regions in all quadrants; the periodic families of $Q_2$ and $Q_4$ are maintained by the libration of resonant angles $\phi_0$, $\phi_3$,$\phi_4$ around 0 and $\phi_1$,$\phi_2$ and $\phi_5$ around $\pi$. For $M = \pi$, there are two periodic families with high $e_p$ in $Q_1$ and $Q_3$. The principal periodic family observed in $Q_3$ in Figure \ref{fig:411} (a) agrees with the fixed point family obtained by \cite{kotoulas2020planar}. While their periodic family extends from $e = 0$ to $e= 0.6$,  we observe it in the range $e = 0.22$ to $e \approx 0.6$. This difference could be due to the limited resolution of our maps. The periodic family near to the x-axis at high $e$ is in agreement with the bifurcation of the eccentric resonant family of the CR3BP reported in \cite{kotoulas2020planar}, but this family could not be computed by these authors due to numerical difficulties. Furthermore, the periodic families in $Q_1$ and $Q_3$ for $M = \pi$ (Figure \ref{fig:411} (b)) also occur with $e_p \approx 0.9$ and were also not identified in \cite{kotoulas2020planar}.

The stability maps for the planetary problem when the 2nd planet has Neptune mass are presented in Figure \ref{fig:412}. For $M = 0$, a periodic family with $e_p \approx 0$ appears in $Q_1$. Both in $Q_1$ and $Q_3$ there is a periodic family for high values of $e$ and $e_p$. The $\phi_3$ libration region of $Q_3$ vanishes with the increase of mass. For $M = \pi$ the fixed point families are similar to the ones observed in the ER3BP map.

The stability maps when the retrograde body has Saturn mass are presented in Figure \ref{fig:413}. For $M = 0$, the periodic families with $e_p \approx 0$, observed in the Neptune case, disappear. The periodic families present in the case of Neptune mass almost disappear in $Q_2$ and vanish in $Q_4$. The $\phi_3$ libration region no longer exists  in $Q_3$, although in $Q_4$ a new region of $\phi_4$ libration around 0 appears. The principal periodic families in $Q_1$ and $Q_3$ decrease in size, while the periodic families with high $e$ and $e_p$ increase a little. For $M = \pi$, the periodic families decreased considerably in size with the increase of mass.  In Figure \ref{fig:41satci} we show the orbital evolution corresponding to the initial condition marked in $Q_4$ of Figure \ref{fig:413} (a); this initial condition is maintained by the single libration of $\phi_4$ around $0$.

In Figure \ref{fig:414}, the stability maps considering the 2nd planet with Jupiter mass are presented. For $M = 0$, we can see that a large region within periodic family is destroyed when both planets have the same mass. This instability occurs because these initial condition are vertically unstable, causing one of the planes to collide with the star. The $\phi_3$ libration region reappears in $Q_3$ and survives the increase of mass in $Q_4$. In $Q_4$, there is a stable region maintained by the libration of $\phi_1$ around $\pi$. Similar to the maps for other 2nd planet masses, there are very few families on map for $M = \pi$; for the Jupiter case, a $\phi_1$ libration region appears in $Q_3$.

\begin{figure}
     \centering
     \begin{subfigure}[b]{0.43\textwidth}
         \centering
         \includegraphics[width=\textwidth]{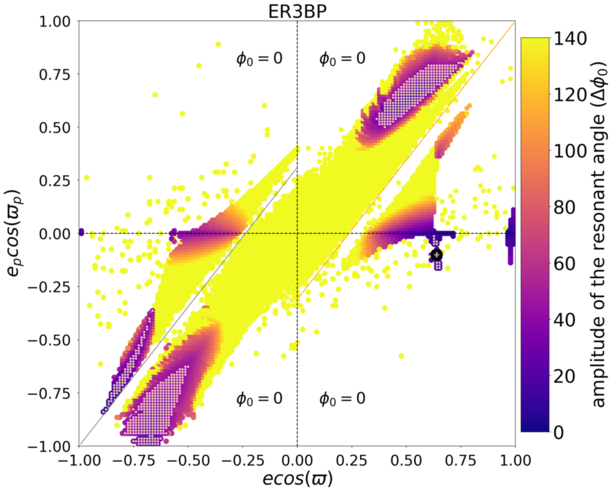}
         \caption{}
         \label{fig:23ER3BP0}
     \end{subfigure}
     \vskip15pt
     \begin{subfigure}[b]{0.43\textwidth}
         \centering
         \includegraphics[width=\textwidth]{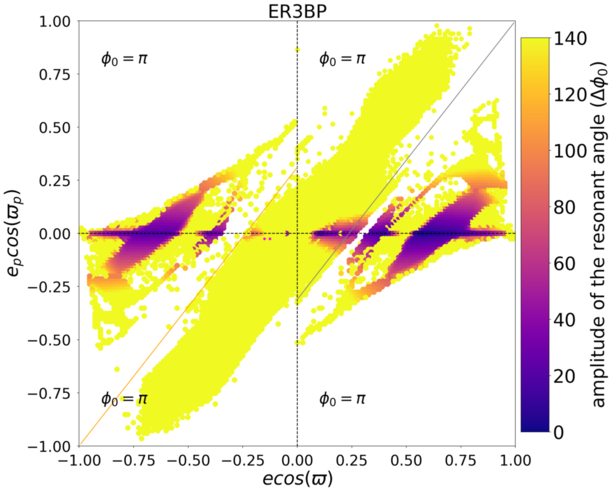}
         \caption{}
         \label{fig:23ER3BP180}
     \end{subfigure}
     \caption{Resonant maps for the 2/-3 resonance in the elliptic restricted three body problem: (a) $M=0$; (b) $M=\pi$. The amplitude of restricted angle ($\phi_{0}$) is represented by the color bar and the overlaying white symbols indicate the fixed point family where all resonant angles librate around a center. The orange and gray lines indicate, respectively, collision at time zero or after a period of the external object.}
     \label{fig:231}
     \end{figure}

\begin{figure}
    \centering
    \includegraphics[width=0.43\textwidth]{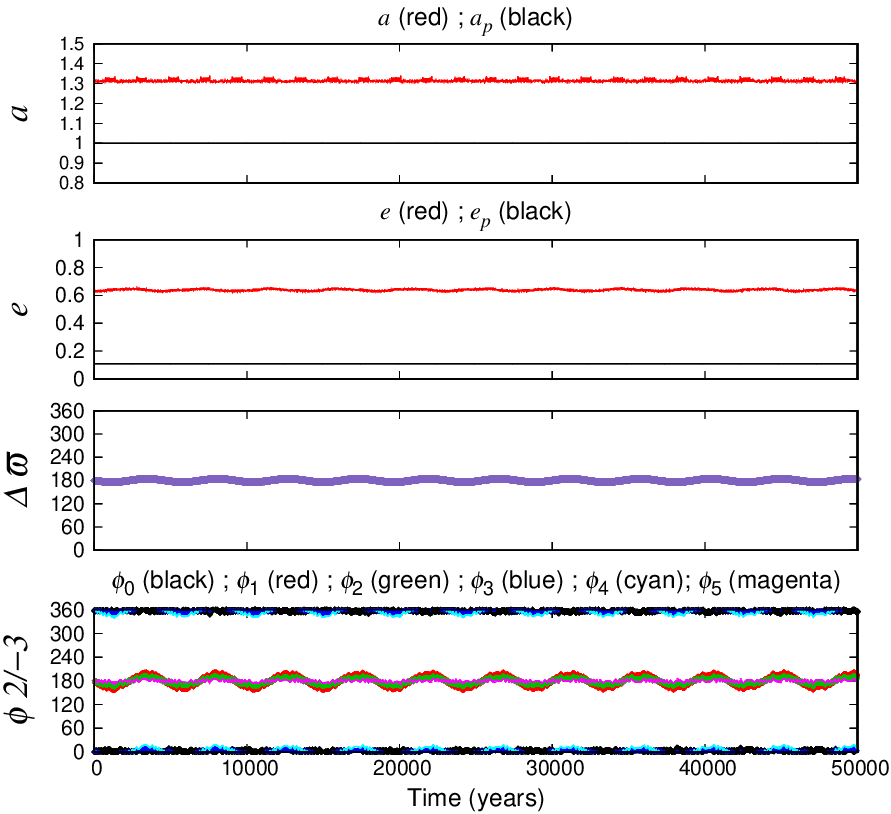}
    \caption{Orbital evolution of the initial condition circled in Figure \ref{fig:231}. The initial condition is $e = 0.63$, $e_p= 0.11$ and corresponds to a fixed point family maintained by the libration of all resonant angles and $\Delta \varpi$. The 1st, 2nd, 3rd and 4th panels show, respectively, the third body's semi-major axis, its eccentricity, the difference $\Delta \varpi$ between the longitudes of pericenter, and the resonant angles $\phi_0$, $\phi_1$, $\phi_2$, $\phi_3$, $\phi_4$ and $\phi_5$.}
\label{fig:23er3bpci}
\end{figure}

\begin{figure}
     \centering
     \begin{subfigure}[b]{0.43\textwidth}
         \centering
         \includegraphics[width=\textwidth]{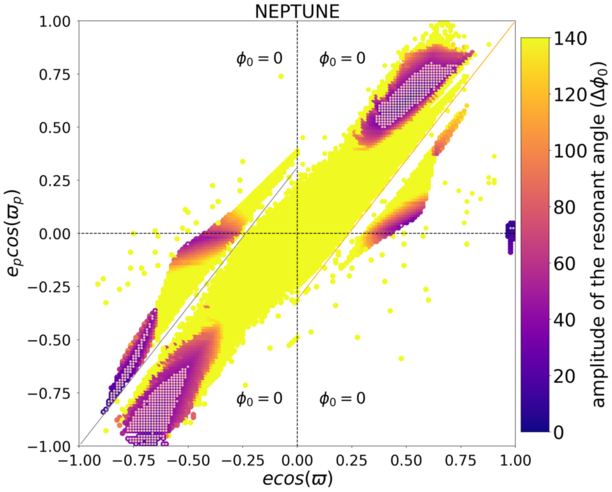}
         \caption{}
         \label{fig:23NEP0}
     \end{subfigure}
     \vskip15pt
     \begin{subfigure}[b]{0.43\textwidth}
         \centering
         \includegraphics[width=\textwidth]{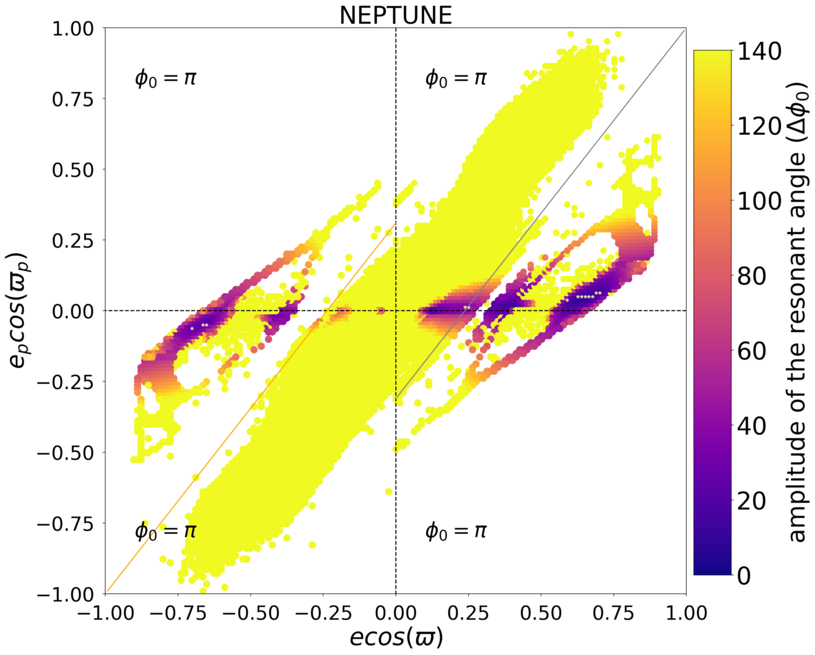}   
         \caption{}
         \label{fig:23NEP180}
     \end{subfigure}
     \caption{Resonant maps for the 2/-3 resonance in the planetary problem when the 2nd planet has Neptune's mass: (a) $M=0$; (b) $M=\pi$. The amplitude of restricted angle ($\phi_{0}$) is represented by the color bar and the overlaying white symbols indicate the fixed point family where all resonant angles librate around a center. The orange and gray lines indicate, respectively, collision at time zero or after a period of the external object.}
     \label{fig:232}
     \end{figure}

\begin{figure}
     \centering
     \begin{subfigure}[b]{0.43\textwidth}
         \centering
         \includegraphics[width=\textwidth]{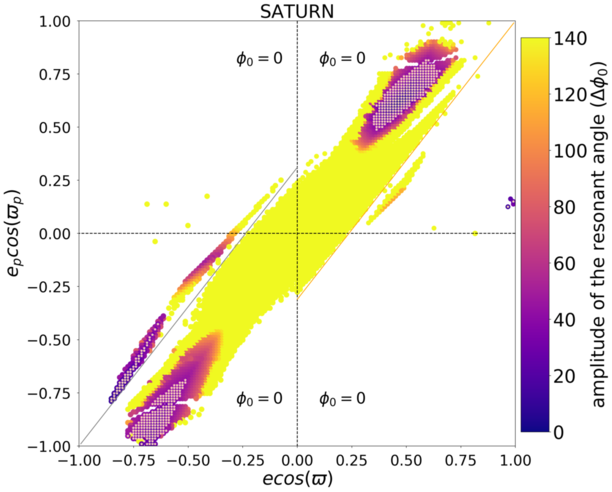}
         \caption{}
         \label{fig:23SAT0}
     \end{subfigure}
     \vskip15pt
     \begin{subfigure}[b]{0.43\textwidth}
         \centering
         \includegraphics[width=\textwidth]{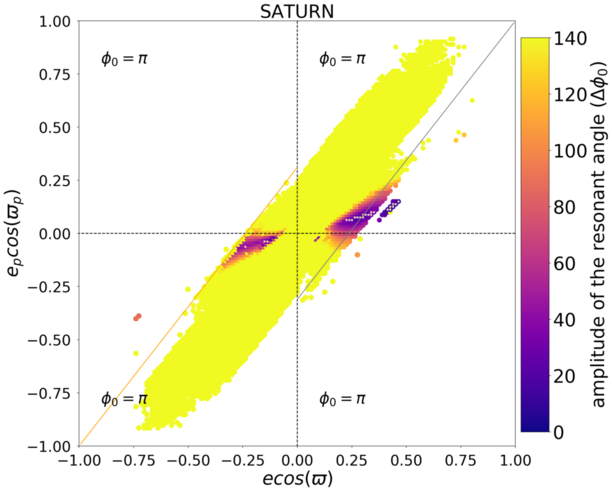}
         \caption{}
         \label{fig:23SAT180}
     \end{subfigure}
     \caption{Resonant maps for the 2/-3 resonance in the planetary problem when the 2nd planet has Saturn's mass: (a) $M=0$; (b) $M=\pi$. The amplitude of restricted angle ($\phi_{0}$) is represented by the color bar and the overlaying white symbols indicate the fixed point family where all resonant angles librate around a center. The orange and gray lines indicate, respectively, collision at time zero or after a period of the external object.}
     \label{fig:233}
     \end{figure}

\begin{figure}
     \centering
     \begin{subfigure}[b]{0.43\textwidth}
         \centering
         \includegraphics[width=\textwidth]{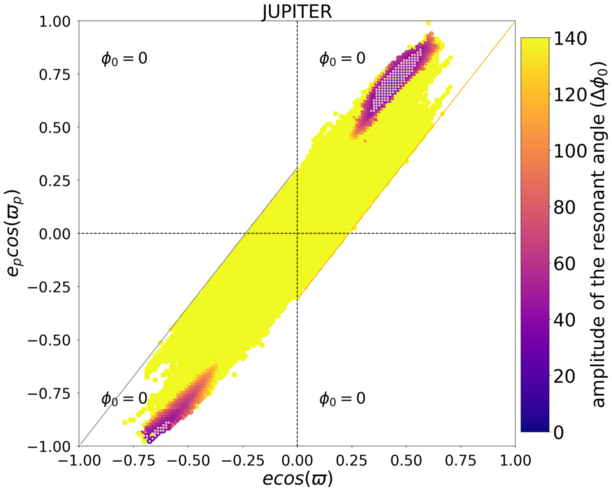}
         \caption{}
         \label{fig:23JUP0}
     \end{subfigure}
     \vskip15pt
     \begin{subfigure}[b]{0.43\textwidth}
         \centering
         \includegraphics[width=\textwidth]{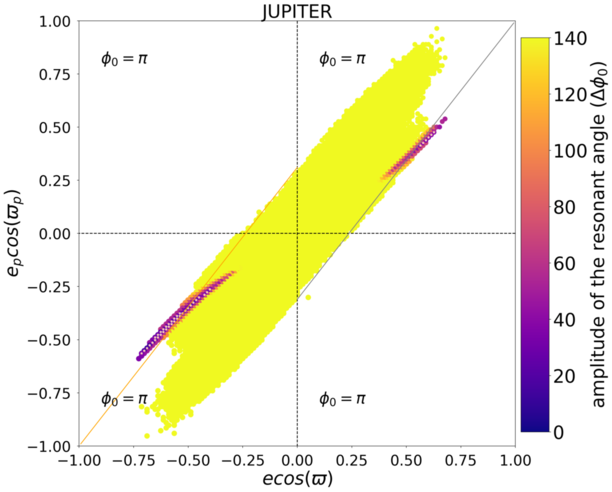}
         \caption{}
         \label{fig:23JUP180}
     \end{subfigure}
     \caption{Resonant maps for the 2/-3 resonance region considering the third body with Jupiter's mass: (a) $M=0$; (b) $M=\pi$. The amplitude of restricted angle ($\phi_{0}$) is represented by the color bar and the overlaying white symbols indicate the fixed point family where all resonant angles librate around a center. The orange and gray lines indicate, respectively, collision at time zero or after a period of the external object.}
     \label{fig:234}
     \end{figure}

\subsection{2/-3 Resonance (Figs. \ref{fig:231}-\ref{fig:234})}
\label{sec23}

The resonant angles analyzed were:

\begin{equation}
    \phi_{0} = -3\lambda - 2\lambda_p + 5\varpi \,\,\, (\text{color bar})
\end{equation}
\begin{equation}
    \phi_1 = -3\lambda - 2\lambda_p + 5\varpi_p   \,\,\, (\text{red})  
\end{equation}
\begin{equation}
   \phi_2 = -3\lambda - 2\lambda_p + 3\varpi_p + 2\varpi   \,\,\, (\text{green})
\end{equation}
\begin{equation}
    \phi_3 = -3\lambda - 2\lambda_p + 2\varpi_p + 3\varpi   \,\,\, (\text{blue})
\end{equation}
\begin{equation}
    \phi_4 = -3\lambda - 2\lambda_p + 4\varpi_p + 1\varpi   \,\,\, (\text{cyan})
\end{equation}
\begin{equation}
    \phi_5 = -3\lambda - 2\lambda_p + 1\varpi_p + 4\varpi   \,\,\, (\text{magenta})
\end{equation}

The stability maps when  we consider a third body with no mass are presented in Figure \ref{fig:231}. The top panel (a) correspond to $M = 0$ which implies $\phi_0 = 0$ in all quadrants, while the bottom panel (b) correspond to $M = \pi$ implying $\phi_0 = \pi$ in all quadrants. The periodic families are present only for $M = 0$, the periodic family in $Q_4$ is maintained by the libration of $\phi_0$, $\phi_3$, $\phi_4$ around 0 and $\phi_1$, $\phi_2$, $\phi_5$ around $\pi$. The other periodic families are maintained by the libration of all resonant angles and $\Delta \varpi$ around 0. For $M = \pi$, there are some regions for low $e_p$ values which are stable due the libration of $\phi_0$ around $\pi$. The two fixed point families in $Q_3$ of Figure \ref{fig:231} (a) are relatively in agreement with the results obtained in \cite{kotoulas2020retrograde} for the ER3BP when the 2nd planet has Neptune's mass. Our $Q_4$ family is slightly different compared to \cite{kotoulas2020retrograde} as in this work the family appears in $Q_2$. These two quadrants correspond to approximately equivalent initial conditions but the the $Q_2$ family did not survive in our numerical integrations. This difference is due to  the different masses  used for the prograde planet. For comparison purposes, we performed the integrations using a prograde planet with the mass of Neptune instead of the mass of Jupiter and we obtained results very similar to those reported in \cite{kotoulas2020retrograde}, the only difference is that we observe fixed point regions with $e_p > 0.6$ and $e \approx 0.45$ in $Q_2$ and $e_p > 0.75$ and $e \approx 0.65$ in $Q_4$, both for $M = 0$. In Figure \ref{fig:23er3bpci} we show the orbital evolution corresponding to the initial condition marked in $Q_4$ of Figure \ref{fig:231} (a).

The stability maps for the planetary problem when the 2nd planet has Neptune mass are presented in Figure \ref{fig:232}. For $M = 0$,  a fixed point family with $e_p \approx 0$ and very high values of retrograde planet eccentricity appears, this family is maintained by the libration of all resonant angles and $\Delta \varpi$ around 0. With the increase of mass of the retrograde body the periodic family in $Q_4$ disappears. For $M = \pi$, the regions maintained by  $\phi_0$ libration decrease and some fixed point initial conditions with $e_p \approx 0$ appear, in these fixed points there is libration of all resonant angles around $\pi$ and $\Delta \varpi$  around $0$. 

The stability maps for the planetary problem when the 2nd planet has Saturn mass are presented in Figure \ref{fig:233}. The regions maintained by $\phi_0$ librations continue to decrease as the mass increases. For both maps, the periodic families with $e_p \approx 0$ are shifted to larger $e_p$ or are destroyed.

The stability maps for the case where both planets have the mass of Jupiter are presented in Figure \ref{fig:234}. For $M = 0$, there are only two periodic families with high $e_p$ values, these are present in $Q_1$ and $Q_3$. The same occurs for $M = \pi$, however these initial conditions are maintained by the resonant angles librations for lower values of $e$ and $e_p$. Remembering that collision lines are approximations derived from the two-body problem, hence due to the high perturbation in this resonance case, this lines are overlapping some of the fixed point families.

\begin{figure}
     \centering
     \begin{subfigure}[b]{0.43\textwidth}
         \centering
         \includegraphics[width=\textwidth]{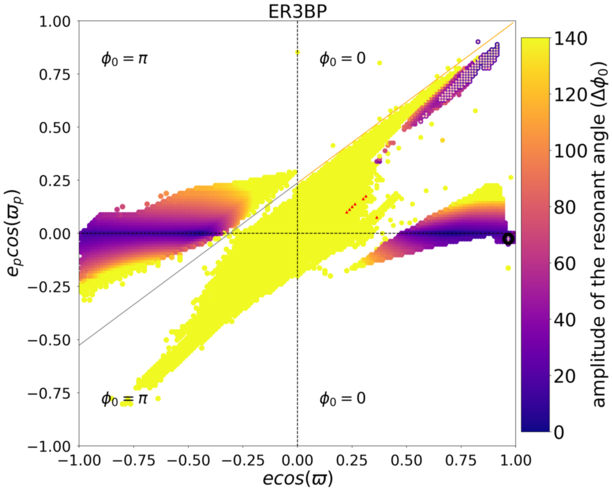}
         \caption{}
         \label{fig:32ER3BP0}
     \end{subfigure}
     \vskip15pt
     \begin{subfigure}[b]{0.43\textwidth}
         \centering
         \includegraphics[width=\textwidth]{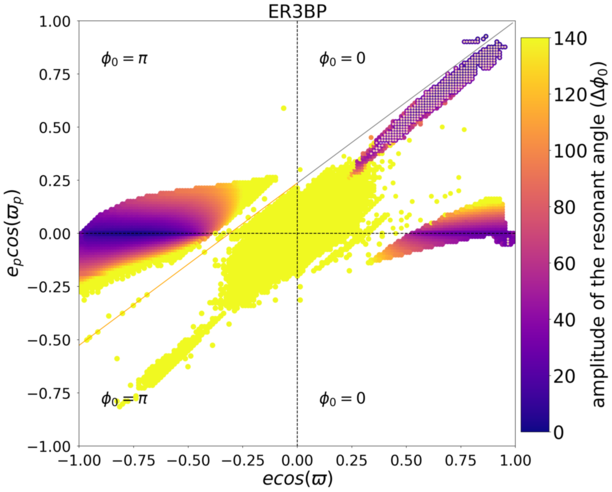}
         \caption{}
         \label{fig:32ER3BP180}
     \end{subfigure}
     \caption{Resonant maps for the 3/-2 resonance in the elliptic restricted three body problem: (a) $M=0$; (b) $M=\pi$. The amplitude of restricted angle ($\phi_{0}$) is represented by the color bar and the overlaying white symbols indicate the fixed point family where all resonant angles librate around a center. The colored symbols indicate libration of a single resonant angle, $\phi_{1}$ (red). The orange and gray lines indicate, respectively, collision at time zero or after a period of the external object.}
     \label{fig:321}
     \end{figure}

\begin{figure}
    \centering
    \includegraphics[width=0.43\textwidth]{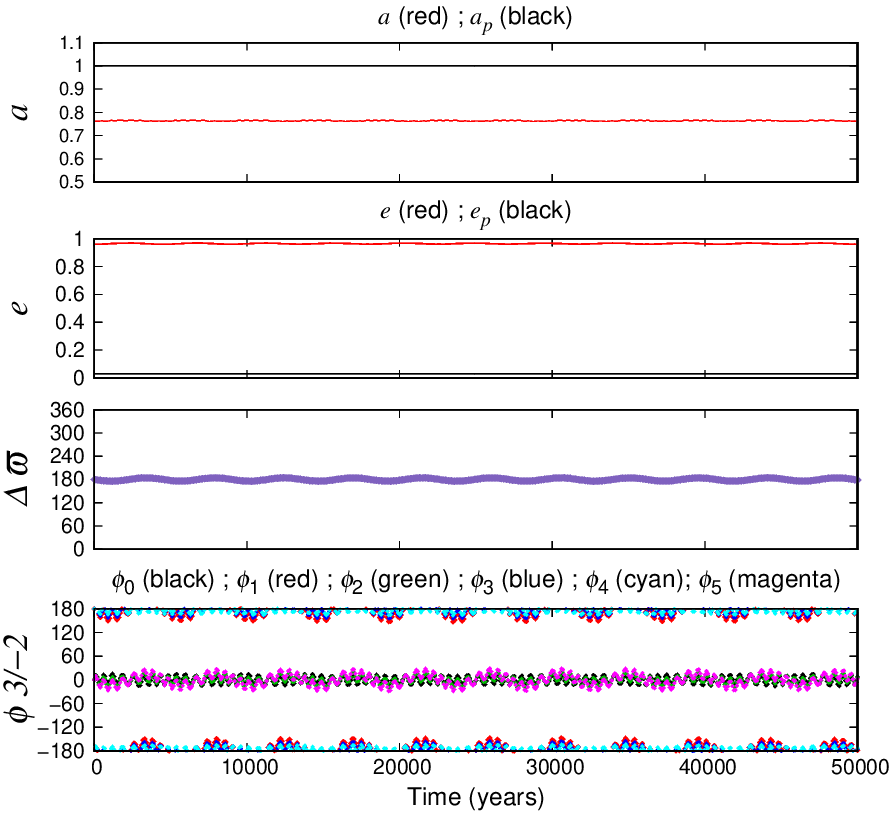}
    \caption{Orbital evolution of the initial condition circled in Figure \ref{fig:321}. The initial condition is $e = 0.96$, $e_p= 0.03$ and corresponds to a fixed point family maintained by the libration of all resonant angles and $\Delta \varpi$. The 1st, 2nd, 3rd and 4th panels show, respectively, the third body's semi-major axis, its eccentricity, the difference $\Delta \varpi$ between the longitudes of pericenter, and the resonant angles $\phi_0$, $\phi_1$, $\phi_2$, $\phi_3$, $\phi_4$ and $\phi_5$.}
\label{fig:32er3bpci}
\end{figure}

\begin{figure}
     \centering
     \begin{subfigure}[b]{0.43\textwidth}
         \centering
         \includegraphics[width=\textwidth]{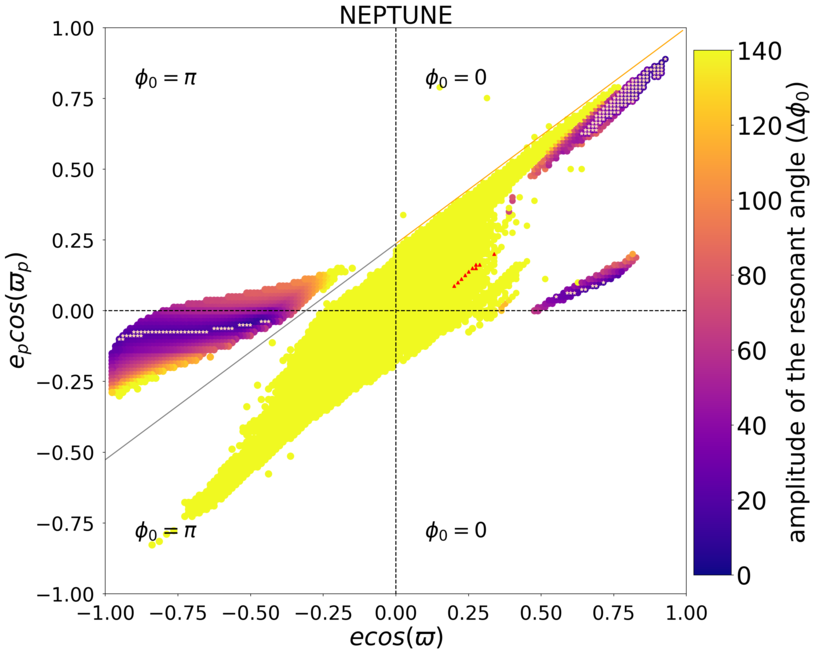}   
         \caption{}
         \label{fig:32NEP0}
     \end{subfigure}
     \vskip15pt
     \begin{subfigure}[b]{0.43\textwidth}
         \centering
         \includegraphics[width=\textwidth]{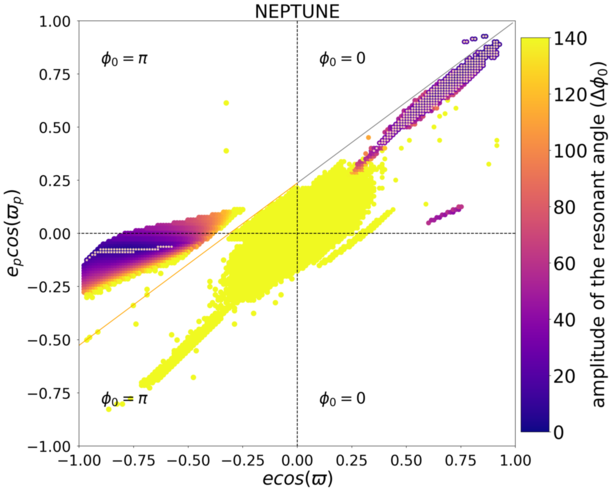}
         \caption{}
         \label{fig:32NEP180}
     \end{subfigure}
     \caption{Resonant maps for the 3/-2 resonance in the planetary problem when the 2nd planet has Neptune's mass: (a) $M=0$; (b) $M=\pi$. The amplitude of restricted angle ($\phi_{0}$) is represented by the color bar and the overlaying white symbols indicate the fixed point family where all resonant angles librate around a center. The colored symbols indicate libration of a single resonant angle, $\phi_{1}$ (red). The orange and gray lines indicate, respectively, collision at time zero or after a period of the external object.}
     \label{fig:322}
     \end{figure}

\begin{figure}
     \centering
     \begin{subfigure}[b]{0.43\textwidth}
         \centering
         \includegraphics[width=\textwidth]{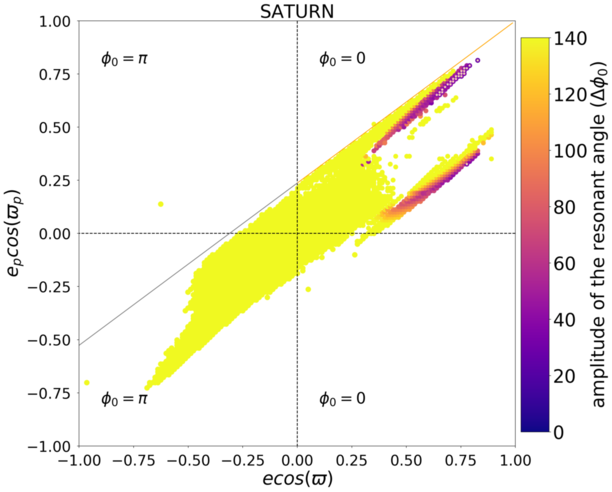}
         \caption{}
         \label{fig:32SAT0}
     \end{subfigure}
     \vskip15pt
     \begin{subfigure}[b]{0.43\textwidth}
         \centering
         \includegraphics[width=\textwidth]{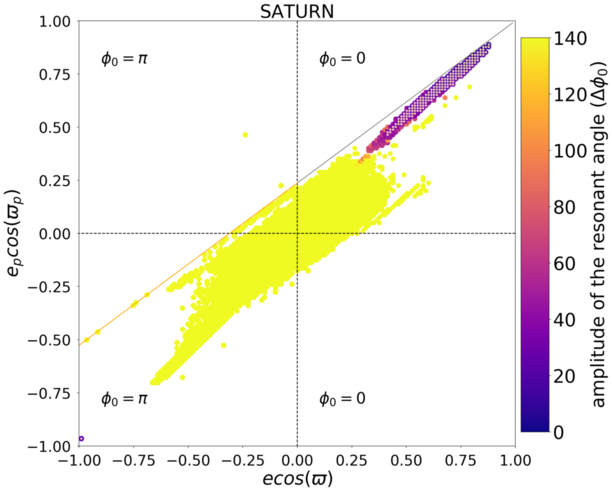}
         \caption{}
         \label{fig:32SAT180}
     \end{subfigure}
     \caption{Resonant maps for the 3/-2 resonance in the planetary problem when the 2nd planet has Saturn's mass: (a) $M=0$; (b) $M=\pi$. The amplitude of restricted angle ($\phi_{0}$) is represented by the color bar and the overlaying white symbols indicate the fixed point family where all resonant angles librate around a center. The orange and gray lines indicate, respectively, collision at time zero or after a period of the external object.}
     \label{fig:323}
     \end{figure}

\begin{figure}
     \centering
     \begin{subfigure}[b]{0.43\textwidth}
         \centering
         \includegraphics[width=\textwidth]{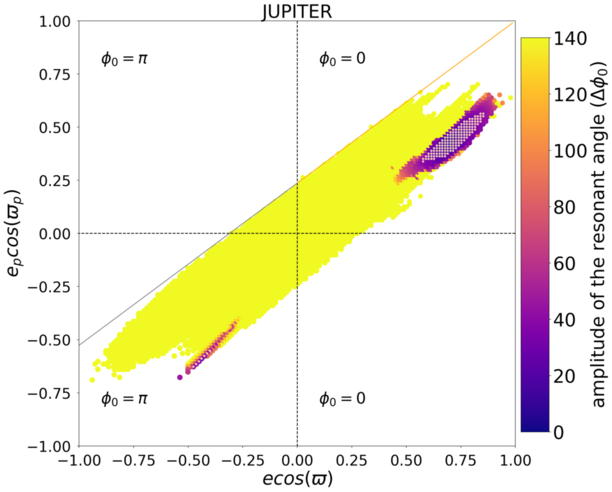}
         \caption{}
         \label{fig:32JUP0}
     \end{subfigure}
     \vskip15pt
     \begin{subfigure}[b]{0.43\textwidth}
         \centering
         \includegraphics[width=\textwidth]{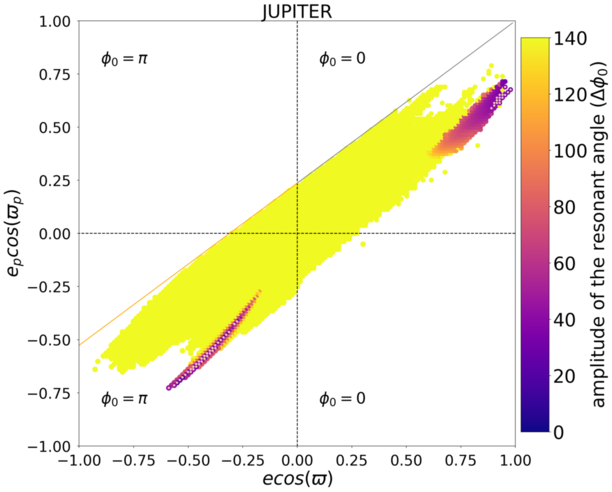}
         \caption{}
         \label{fig:32JUP180}
     \end{subfigure}
     \caption{Resonant maps for the 3/-2 resonance region considering the third body with Jupiter's mass: (a) $M=0$; (b) $M=\pi$. The amplitude of restricted angle ($\phi_{0}$) is represented by the color bar and the overlaying white symbols indicate the fixed point family where all resonant angles librate around a center. The orange and gray lines indicate, respectively, collision at time zero or after a period of the external object.}
     \label{fig:324}
     \end{figure}

\subsection{3/-2 Resonance (Figs. \ref{fig:321}-\ref{fig:324})} 

The resonant angles analyzed were:

\begin{equation}
    \phi_{0} = -2\lambda - 3\lambda_p + 5\varpi \,\,\, (\text{color bar})
\end{equation}
\begin{equation}
    \phi_1 = -2\lambda - 3\lambda_p + 5\varpi_p   \,\,\, (\text{red})  
\end{equation}
\begin{equation}
   \phi_2 = -2\lambda - 3\lambda_p + 2\varpi_p + 3\varpi   \,\,\, (\text{green})
\end{equation}
\begin{equation}
    \phi_3 = -2\lambda - 3\lambda_p + 3\varpi_p + 2\varpi   \,\,\, (\text{blue})
\end{equation}
\begin{equation}
    \phi_4 = -2\lambda - 3\lambda_p + \varpi_p + 4\varpi   \,\,\, (\text{cyan})
\end{equation}
\begin{equation}
    \phi_5 = -2\lambda - 3\lambda_p + 4\varpi_p + \varpi   \,\,\, (\text{magenta})
\end{equation}

In Figure \ref{fig:321} we present the maps for the ER3BP. The top and bottom panels, represented by (a) and (b), were obtained respectively for $M = 0$ and $M = \pi$, which implies $\phi_0 = 0$ in $Q_1$, $Q_4$ and $\phi_0 = \pi$ in $Q_2$, $Q_3$ in both maps. The symmetry for this resonance occurs between equivalent quadrants for $M = 0$ and $M = \pi$. There are periodic families in $Q_1$ and $Q_4$ for both asteroid mean anomaly values, the $Q_4$ family is maintained by the libration of $\phi_0, \phi_2, \phi_5$ around 0 and $\phi_1, \phi_3, \phi_4, \Delta \varpi$ around $\pi$. For $M = 0$, there are some initial conditions maintained by the single libration of $\phi_1$ around 0. The principal periodic family  observed in $Q_1$ in both maps of Figure \ref{fig:321} is partially in agreement with the fixed point family obtained by \cite{kotoulas2020planar}, again the difference occurs in relation to the range of eccentricity which extends to $e=0$ in the latter work. Again, this difference is probably explained by the limited resolution of our maps.  The small periodic family with $e = 0.96$ and $e_p = 0.03$ observed in $Q_4$ also agrees with the results presented in \cite{kotoulas2020planar}. In Figure \ref{fig:32er3bpci} we show the orbital variation corresponding to the initial condition marked in $Q_4$ of Figure \ref{fig:321}.

The stability maps for the planetary problem when the 2nd planet has Neptune mass are presented in Figure \ref{fig:322}. For $M = 0$, there are new periodic families with low values of prograde planet eccentricity in $Q_1$ and $Q_3$. The $\phi_1$ libration region is still present. For $M = \pi$, a periodic family with low values of $e_p$ also appears in $Q_3$. The periodic family with $e = 0.96$ present in $Q_4$ of ER3BP map vanishes with the increase in retrograde body mass.

The stability maps for the planetary problem when the 2nd planet has Saturn mass are presented in Figure \ref{fig:323}. In general, for both values of mean anomaly, the family of $Q_3$ disappears. For $M = 0$, the periodic family with low values of $e_p$ observed in $Q_1$ in the Neptune case mostly disappered, there is only an initial condition with semi-amplitude of resonant angles less than $45^\circ$. The $\phi_1$ libration region does not exist for this case.

In Figure \ref{fig:324}, the stability maps considering the retrograde planet with Jupiter mass are presented. In both maps, there is a periodic family in $Q_3$, maintained by libration of all resonant angles and $\Delta \varpi$ around $\pi$. When both planets have the mass of Jupiter, the two principal families observed in $Q_1$ exist for lower values of the prograde planet eccentricity.

\section{Conclusion}

In this article we showed that there are stable configurations for the fourth and fifth order retrograde resonances when we consider a system composed by a solar mass star, a Jupiter mass planet with prograde motion and a retrograde planet with either zero mass (ER3BP), or non-zero mass equal to Neptune, Saturn or Jupiter. As we increase the mass of the retrograde planet, we observe an expressive change in the resonant phase space.

In general, when we increase the mass of the 2nd planet to the mass of Neptune some additional fixed point families close to the x-axis ($e_p \approx 0$) appear. As we continue to increase the mass of the retrograde planet, these families move away from the x-axis and the fixed point region become less predominant in the phase space, that is, when both planets have the mass of Jupiter, there are fewer initial conditions for families of fixed points than for other masses. This happens for all retrograde resonances of fourth and fifth order. Except for the 2/-3 and 3/-2 resonances, we observe that a considerable region of the main fixed point family becomes vertically unstable when both planets have the same mass. This instability is not caused by close encounters but rather by the increase/decrease of the inclination of the prograde/retrograde body  until one of the two collides with the star.

We observe some differences between our numerical results and the periodic families reported in \cite{kotoulas2020retrograde, kotoulas2020planar}. In the case of the 3/-1 resonance, we observe a new fixed point family and a family at high eccentricity $e_p$. For the 4/-1 resonance, we also obtain families at high $e_p$ and we recover a fixed point family close to the x-axis at high $e$ which is in agreement with the bifurcation from the resonant eccentric family of the CR3BP reported in \cite{kotoulas2020planar} but that could not be computed by these authors due to numerical difficulties. For the 4/-1 and 3/-2 resonances, the fixed point families which occur due to bifurcation of the circular family of the CR3BP, and are reported in \cite{kotoulas2020planar}, do not extend to $e=0$ and $e_p=0$ in our work. This is likely due to the different methods used. While  \cite{kotoulas2020planar} compute the periodic families using the method of continuation from the CR3BP, we obtain information about these families by computing stability maps where we observe the quasi-periodic regions around the stable periodic families. When these quasi-periodic regions are small, our map resolution may not be enough to identify the families. However, the information about the extent of the quasi-periodic regions is important and therefore our work is complementary to the work in \cite{kotoulas2020planar}.
Furthermore, for the 2/-3 resonance, we observe an expected disagreement due to the difference of the prograde planet mass used in this article and in \cite{kotoulas2020retrograde}. With additional simulations using the mass of Neptune for the prograde planet, we obtain, except for families at high $e_p$, results in agreement with \cite{kotoulas2020retrograde}.

As proposed by \cite{gayon2008retrograde, gayon2009fitting}, resonant exo-planetary systems with counter revolving motion may exist. In some cases, the fitting of radial velocity curves considering such retrograde configurations is better than for prograde configurations. Our results indicate which stable retrograde configurations are possible in a system with planets in a fourth or fifth-order resonance. Therefore, our work may be used as a guide for searching  such systems.

\section*{Acknowledgements}

This work was funded by the grants FAPESP/2021/11982-5 and FAPESP/2022/08716-4 of São Paulo Research Foundation. The authors acknowledge support from Coordenação de Aperfeiçoamento de Pessoal de Nível Superior – Brasil (CAPES) – Finance Code 001 (88887.675709/2022-00). The computational resources were supplied by the Center for Scientific Computing (NCC/GridUNESP) of the São Paulo State University (UNESP). 

\section*{Data Availability}

The data underlying this paper will be shared on reasonable request
to the corresponding author.



\bibliographystyle{mnras}
\bibliography{example} 



\appendix

\bsp	
\label{lastpage}
\end{document}